\PassOptionsToPackage{unicode}{hyperref}
\PassOptionsToPackage{hyphens}{url}
\PassOptionsToPackage{dvipsnames,svgnames,x11names}{xcolor}
\documentclass[
]{article}
\usepackage{xcolor}
\usepackage[margin=1.1in]{geometry}
\usepackage{amsmath,amssymb}
\usepackage{iftex}
\ifPDFTeX
  \usepackage[T1]{fontenc}
  \usepackage[utf8]{inputenc}
  \usepackage{textcomp} 
\else 
  \usepackage{unicode-math} 
  \defaultfontfeatures{Scale=MatchLowercase}
  \defaultfontfeatures[\rmfamily]{Ligatures=TeX,Scale=1}
\fi
\usepackage{lmodern}
\ifPDFTeX\else
\fi
\IfFileExists{upquote.sty}{\usepackage{upquote}}{}
\IfFileExists{microtype.sty}{
  \usepackage[]{microtype}
  \UseMicrotypeSet[protrusion]{basicmath} 
}{}
\makeatletter
\@ifundefined{KOMAClassName}{
  \IfFileExists{parskip.sty}{%
    \usepackage{parskip}
  }{
    \setlength{\parindent}{0pt}
    \setlength{\parskip}{6pt plus 2pt minus 1pt}}
}{
  \KOMAoptions{parskip=half}}
\makeatother
\makeatletter
\ifx\paragraph\undefined\else
  \let\oldparagraph\paragraph
  \renewcommand{\paragraph}{
    \@ifstar
      \xxxParagraphStar
      \xxxParagraphNoStar
  }
  \newcommand{\xxxParagraphStar}[1]{\oldparagraph*{#1}\mbox{}}
  \newcommand{\xxxParagraphNoStar}[1]{\oldparagraph{#1}\mbox{}}
\fi
\ifx\subparagraph\undefined\else
  \let\oldsubparagraph\subparagraph
  \renewcommand{\subparagraph}{
    \@ifstar
      \xxxSubParagraphStar
      \xxxSubParagraphNoStar
  }
  \newcommand{\xxxSubParagraphStar}[1]{\oldsubparagraph*{#1}\mbox{}}
  \newcommand{\xxxSubParagraphNoStar}[1]{\oldsubparagraph{#1}\mbox{}}
\fi
\makeatother

\usepackage{longtable,booktabs,array}
\usepackage{calc} 
\usepackage{etoolbox}
\makeatletter
\patchcmd\longtable{\par}{\if@noskipsec\mbox{}\fi\par}{}{}
\makeatother
\IfFileExists{footnotehyper.sty}{\usepackage{footnotehyper}}{\usepackage{footnote}}
\makesavenoteenv{longtable}
\usepackage{graphicx}
\makeatletter
\newsavebox\pandoc@box
\newcommand*\pandocbounded[1]{
  \sbox\pandoc@box{#1}%
  \Gscale@div\@tempa{\textheight}{\dimexpr\ht\pandoc@box+\dp\pandoc@box\relax}%
  \Gscale@div\@tempb{\linewidth}{\wd\pandoc@box}%
  \ifdim\@tempb\p@<\@tempa\p@\let\@tempa\@tempb\fi
  \ifdim\@tempa\p@<\p@\scalebox{\@tempa}{\usebox\pandoc@box}%
  \else\usebox{\pandoc@box}%
  \fi%
}
\def\fps@figure{htbp}
\makeatother

\providecommand{\tightlist}{%
  \setlength{\itemsep}{0pt}\setlength{\parskip}{0pt}}

\usepackage[backend=biber,style=authoryear]{biblatex}
\makeatletter
\@ifpackageloaded{caption}{}{\usepackage{caption}}
\AtBeginDocument{%
\ifdefined\contentsname
  \renewcommand*\contentsname{Table of contents}
\else
  \newcommand\contentsname{Table of contents}
\fi
\ifdefined\listfigurename
  \renewcommand*\listfigurename{List of Figures}
\else
  \newcommand\listfigurename{List of Figures}
\fi
\ifdefined\listtablename
  \renewcommand*\listtablename{List of Tables}
\else
  \newcommand\listtablename{List of Tables}
\fi
\ifdefined\figurename
  \renewcommand*\figurename{Figure}
\else
  \newcommand\figurename{Figure}
\fi
\ifdefined\tablename
  \renewcommand*\tablename{Table}
\else
  \newcommand\tablename{Table}
\fi
}
\@ifpackageloaded{float}{}{\usepackage{float}}
\floatstyle{ruled}
\@ifundefined{c@chapter}{\newfloat{codelisting}{h}{lop}}{\newfloat{codelisting}{h}{lop}[chapter]}
\floatname{codelisting}{Listing}

\makeatother
\makeatletter
\@ifpackageloaded{caption}{}{\usepackage{caption}}
\@ifpackageloaded{subcaption}{}{\usepackage{subcaption}}
\makeatother
\usepackage{bookmark}
\IfFileExists{xurl.sty}{\usepackage{xurl}}{} 
\makeatletter
\@ifundefined{xmpquote}{\newcommand{\xmpquote}[1]{#1}}{}
\makeatother
\hypersetup{
  pdftitle={Leveraging Multi-Model Land-Use Emissions Reveals a Rising Airborne Fraction of Atmospheric CO\_2 Over the Past Six Decades},
  pdfauthor={J. Eduardo Vera-Valdés},
  pdfkeywords={\xmpquote{airborne fraction}, \xmpquote{carbon
cycle}, \xmpquote{remaining carbon budget}, \xmpquote{land-use change
emissions}, \xmpquote{carbon sinks}, \xmpquote{trend estimation}},
  colorlinks=true,
  linkcolor={blue},
  filecolor={Maroon},
  citecolor={Blue},
  urlcolor={Blue},
  pdfcreator={LaTeX via pandoc}}

\title{\textbf{Leveraging Multi-Model Land-Use Emissions Reveals a
Rising Airborne Fraction of Atmospheric \(CO_2\) Over the Past Six
Decades}}
\author{%
\begin{tabular}[t]{@{}l@{}}
\textbf{J. Eduardo Vera-Valdés} \\
Aalborg University, Department of Mathematical Sciences \\
CoRE, Center for Research in Energy: Economics and Markets \\
\href{mailto:eduardo@math.aau.dk}{eduardo@math.aau.dk} \\
\href{https://orcid.org/0000-0002-0337-8055}{ORCID: 0000-0002-0337-8055}
\end{tabular}}
\date{}
\begin{document}
\maketitle
\begin{abstract}
The airborne fraction, the share of anthropogenic \(CO_2\) emissions
that remains in the atmosphere rather than being taken up by land and
ocean, is a compact diagnostic of sink efficiency and a direct input to
the remaining carbon budgets behind climate targets. Here we show that
it has risen over the observational record, from \(0.32\)--\(0.40\) in
\(1959\) to \(0.47\)--\(0.50\) by \(2024\) (ranges across estimators),
above the \(0.44\) commonly reported for recent decades. Applied to IPCC
AR6 assessments, this rise reduces the remaining carbon budget by
\(23\)--\(48\) \(GtCO_2\) for a given temperature target, up to roughly
one year of current global fossil emissions. The trend has been
difficult to detect because two features of the standard analysis
obscure it. First, three high-leverage years following the \(1991\)
Pinatubo eruption dominate the volcanic term of the natural-variability
adjustment and attenuate the estimated trend; smoothing those years, or
using a volcanically robust \(ENSO\) index, removes the artifact.
Second, collapsing land-use and land-cover change emissions to a single
series discards the disagreement among models that carries the
information needed to resolve a small trend. Propagating the dispersion
of \(69\) land-use series together with the reported fossil-emission
uncertainty into delta-method weights, weighted and generalized least
squares recover a positive and significant trend, corroborated by a
mixed-effects model and robust across a grid of \(49\)
natural-variability specifications that maps which treatment choices
reproduce which published result.
\end{abstract}

\section*{Key Points}\label{key-points}
\addcontentsline{toc}{section}{Key Points}

\begin{itemize}
\tightlist
\item
  Pinatubo years and one collapsed land-use series have masked a rising
  airborne fraction, hiding it from standard trend estimates
\item
  Weighting 69 land-use series by their dispersion and fossil-emission
  uncertainty recovers a positive, significant trend across estimators
\item
  The airborne fraction rose from about 0.32-0.40 in 1959 to 0.47-0.50
  by 2024, cutting the remaining carbon budget by 23-48 GtCO2
\end{itemize}

\section*{Plain Language Summary}\label{plain-language-summary}
\addcontentsline{toc}{section}{Plain Language Summary}

About half of the carbon dioxide that people release each year stays in
the atmosphere; the rest is taken up by plants, soils, and the ocean.
The share that stays airborne measures how well those sinks are keeping
pace with our emissions, and it is used to work out how much carbon can
still be emitted before a temperature target is passed. Researchers have
disagreed for two decades over whether that share is growing, reaching
opposite answers from almost the same data. We show that it has grown,
from roughly a third of emissions in 1959 to about half today, and that
the disagreement came from two things. The first is the 1991 eruption of
Mount Pinatubo: its three unusual years carry so much weight in the
statistical correction for natural climate variability that they hide
the long-term change. The second is that estimates of emissions from
deforestation and other land-use change differ widely between models,
and using only one of them throws that information away. Using 69
land-use estimates together, and handling the eruption carefully,
reveals a clear upward trend, which implies the remaining carbon budget
is smaller than currently assumed.

\section{Introduction}\label{introduction}

The remaining carbon budget, the amount of anthropogenic carbon dioxide
(\(CO_2\)) that can still be emitted while meeting a given temperature
goal, depends critically on how efficiently the land--ocean system
continues to absorb \(CO_2\). A compact diagnostic of that efficiency is
the airborne fraction (\(AF\)), the share of anthropogenic \(CO_2\)
emissions that remains in the atmosphere. The \(AF\) has been estimated
to be around \(0.44\) in recent decades, but whether it has been rising
has been contested, with studies reaching opposite conclusions from
essentially the same annual carbon-budget data
\autocite{Canadell2007,Raupach2007,Knorr2009,Ballantyne2012,LeQuere2009,bennedsenEvidenceTrendCO22023,bennedsenRegressionbasedApproachCO22024,bennettQuantificationAirborneFraction2024,veravaldes2025robustestimationco2}.

In its classical form, \(AF\) is a yearly ratio of atmospheric growth to
total anthropogenic emissions, computed as the sum of fossil fuel
emissions and land-use and land-cover change emissions as:

\begin{equation}\protect\phantomsection\label{eq-af-def}{
AF_t = \frac{G_t}{FF_t + LULC_t},
}\end{equation}

where \(G_t\) is the annual atmospheric \(CO_2\) growth, \(FF_t\) is
fossil fuel emissions excluding carbonation, and \(LULC_t\) is land-use
and land-cover change emissions.

Even though the data are well established, the literature has not
converged on whether \(AF\) is rising. We show that the disagreement
traces to two distinct statistical problems, which we resolve in this
study. The first statistical problem is an identification one:
atmospheric \(CO_2\) growth is affected by natural variability, so the
standard approach filters it by regressing growth on volcanic and El
Niño--Southern Oscillation (\(ENSO\)) indices. The \(1991\)--\(1993\)
Mount Pinatubo eruption produced aerosol values much larger than any
other event in the record, so three Pinatubo years dominate the volcanic
coefficient and, if left untreated, this single episode can drive the
significance of the estimated trend. The second statistical problem is a
power one: the sample is short, and \(AF\) inference hinges on
denominator uncertainty, which depends on fossil fuel emissions and
land-use and land-cover change (\(LULC\)) uncertainty. There are several
competing \(LULC\) models, and the disagreement among them can be
substantial. Ordinary least squares (\(OLS\)) estimates built on
different \(LULC\) series can therefore meaningfully disagree, and an
\(OLS\) estimate built on one collapsed \(LULC\) series discards the
disagreement between models and lacks the power to resolve a small
trend.

We address the identification problem by showing that the Pinatubo years
are high-leverage observations: smoothing the high values, or using the
Relative Oceanic Niño Index (\(RONI\)) \autocite{LHeureux2024RONI},
which is less sensitive to the choice of climatological base period,
recovers a positive and significant trend across estimators and
natural-variability indices. We address the power problem by integrating
the disagreement among \(69\) \(LULC\) measurement series derived from
the Global Carbon Budget \(2025\) \autocite{Friedlingstein2025},
together with the reported fossil-emission uncertainty, into uncertainty
weights for a single airborne-fraction series, estimated by weighted
least squares (\(WLS\)) and generalized least squares (\(GLS\)); a
mixed-effects model (\(MEM\)) across the ensemble corroborates the
result. Together these estimators give a positive and significant \(AF\)
trend across specifications: the \(AF\) has risen to about
\(0.47\)--\(0.50\) by \(2024\), up from roughly \(0.32\)--\(0.40\) in
\(1959\) (ranges across estimators). This implies a reduction of about
\(23\)--\(48\) \(GtCO_2\) in the remaining carbon budget for a given
temperature target, or up to roughly one year of current global fossil
emissions.

\section{Results}\label{results}

\subsection{Data and estimation
strategy}\label{data-and-estimation-strategy}

\begin{figure}

\centering{

\pandocbounded{\includegraphics[keepaspectratio]{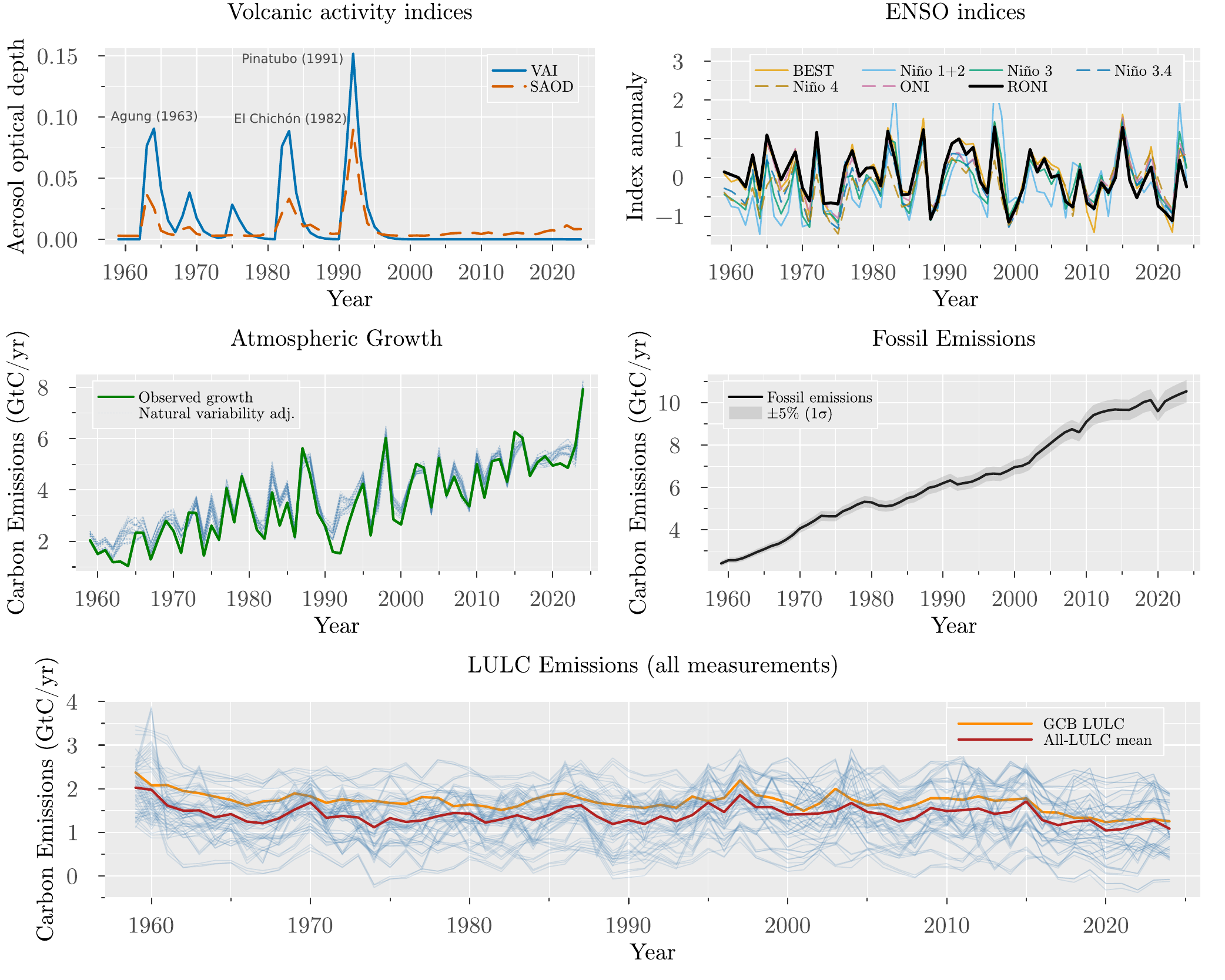}}

}

\caption{\label{fig-data}\emph{Topleft:} Volcanic activity indices. The
three major eruptions in the sample are annotated. \emph{Topright:} The
seven \(ENSO\) indices. \emph{Centre left:} Atmospheric growth raw and
natural-variability adjusted using all combinations of volcanic and
\(ENSO\) indices. \emph{Centre right:} Fossil emissions. \emph{Bottom:}
Land-use and land-cover change (\(LULC\)) panel. Also shown the \(GCB\)
\(LULC\) column and the all-\(LULC\) cross-series mean. All data in the
period \(1959\)--\(2024\).}

\end{figure}

We use annual Global Carbon Budget (\(GCB\)) \(2025\) data
\autocite{Friedlingstein2025}: atmospheric \(CO_2\) growth, fossil
emissions excluding carbonation, and a panel of \(69\) \(LULC\)
measurement series built from the three bookkeeping models (\(BLUE\),
\(OSCAR\), \(LUCE\)) and peat-augmented process-based model combinations
(Methods). Natural variability is controlled by two volcanic indices and
seven \(ENSO\) indices. The data are shown in Figure~\ref{fig-data}.

The analysis proceeds in two stages (Methods). A first-stage regression
filters atmospheric growth for natural variability using an \(ENSO\)
index and a volcanic index. Our preferred specification uses the
\(SAOD\) and \(RONI\), which adjusts Niño 3.4 anomalies by the
contemporaneous tropical sea-surface-temperature anomaly to yield an
\(ENSO\) index less sensitive to the choice of climatological base
period and more comparable across the historical record. A second stage
estimates the linear trend of the resulting airborne-fraction series by
\(WLS\) and \(GLS\), weighting each year by the inverse of a
delta-method variance that propagates the disagreement among the \(69\)
\(LULC\) series, together with the Global Carbon Budget's reported
\(\pm5\%\) fossil-emission uncertainty, into the trend estimate. A
\(MEM\) fitted to the full panel serves as a corroborating cross-check.

\subsection{Pinatubo leverage in the volcanic
adjustment}\label{pinatubo-leverage-in-the-volcanic-adjustment}

Table~\ref{tbl-growth-reg} reports the \(OLS\), \(WLS\), and \(GLS\)
estimates of the growth regression (Methods). The total-emissions and
\(RONI\) coefficients are positive and strongly significant across
estimators. The \(SAOD\) coefficient is large, negative, and significant
(\(\approx -29.7\), \(p<0.001\)), but this coefficient is almost
entirely identified by the three Pinatubo years. When those years are
smoothed to the full-sample mean (Methods), the \(SAOD\) coefficient
falls in magnitude to about \(-11.5\) and loses significance
(\(p\approx 0.32\)), and to about \(-7.3\) (\(p\approx 0.50\)) under the
non-Pinatubo mean. In contrast, the total-emissions coefficient is
essentially unchanged and \(RONI\) remains strongly significant (last
two columns of Table~\ref{tbl-growth-reg}). The Pinatubo years are
therefore high-leverage observations that disproportionately influence
the estimated coefficients, and the volcanic term functions as a
Pinatubo correction rather than a general volcanic adjustment. This high
leverage could explain the divergence between studies, and motivates the
smoothed specifications reported in Table~\ref{tbl-ladder}.

\begin{longtable}[]{@{}
  >{\raggedright\arraybackslash}p{(\linewidth - 10\tabcolsep) * \real{0.2857}}
  >{\raggedleft\arraybackslash}p{(\linewidth - 10\tabcolsep) * \real{0.1143}}
  >{\raggedleft\arraybackslash}p{(\linewidth - 10\tabcolsep) * \real{0.1143}}
  >{\raggedleft\arraybackslash}p{(\linewidth - 10\tabcolsep) * \real{0.1429}}
  >{\raggedleft\arraybackslash}p{(\linewidth - 10\tabcolsep) * \real{0.1714}}
  >{\raggedleft\arraybackslash}p{(\linewidth - 10\tabcolsep) * \real{0.1714}}@{}}
\caption{Preferred growth specification with natural-variability
controls (\(RONI\) lagged one year and contemporaneous \(SAOD\)). The
first three columns use \(SAOD\) as observed (\(OLS\), \(WLS\),
\(GLS\)); the last two columns report the \(GLS\) estimates when the
\(1991\)--\(1993\) Pinatubo \(SAOD\) values are replaced by the
full-series mean or the non-Pinatubo mean. Standard errors are
HAC-robust.}\label{tbl-growth-reg}\tabularnewline
\toprule\noalign{}
\begin{minipage}[b]{\linewidth}\raggedright
Growth equation
\end{minipage} & \begin{minipage}[b]{\linewidth}\raggedleft
\(OLS\)
\end{minipage} & \begin{minipage}[b]{\linewidth}\raggedleft
\(WLS\)
\end{minipage} & \begin{minipage}[b]{\linewidth}\raggedleft
\(GLS\)
\end{minipage} & \begin{minipage}[b]{\linewidth}\raggedleft
\(GLS\) (\(SAOD^*\) full)
\end{minipage} & \begin{minipage}[b]{\linewidth}\raggedleft
\(GLS\) (\(SAOD^*\) excl.)
\end{minipage} \\
\midrule\noalign{}
\endfirsthead
\toprule\noalign{}
\begin{minipage}[b]{\linewidth}\raggedright
Growth equation
\end{minipage} & \begin{minipage}[b]{\linewidth}\raggedleft
\(OLS\)
\end{minipage} & \begin{minipage}[b]{\linewidth}\raggedleft
\(WLS\)
\end{minipage} & \begin{minipage}[b]{\linewidth}\raggedleft
\(GLS\)
\end{minipage} & \begin{minipage}[b]{\linewidth}\raggedleft
\(GLS\) (\(SAOD^*\) full)
\end{minipage} & \begin{minipage}[b]{\linewidth}\raggedleft
\(GLS\) (\(SAOD^*\) excl.)
\end{minipage} \\
\midrule\noalign{}
\endhead
\bottomrule\noalign{}
\endlastfoot
Intercept & -0.5247 & -0.5238 & -0.5096 & -0.7946 & -0.8402 \\
SE (Intercept) & 0.3266 & 0.3247 & 0.3074 & 0.4060 & 0.4048 \\
\(p\)-value (Intercept) & 0.1082 & 0.1067 & 0.0973 & 0.0503 & 0.0379 \\
Total emissions & 0.5450 & 0.5454 & 0.5438 & 0.5568 & 0.5588 \\
SE (Total emissions) & 0.0411 & 0.0409 & 0.0395 & 0.0474 & 0.0471 \\
\(p\)-value (Total emissions) & \textless{} 0.001 & \textless{} 0.001 &
\textless{} 0.001 & \textless{} 0.001 & \textless{} 0.001 \\
\(SAOD\) & -29.3558 & -29.3990 & -29.7478 & -11.5258 & -7.2979 \\
SE (\(SAOD\)) & 4.9699 & 4.9723 & 4.8488 & 11.5799 & 10.7501 \\
\(p\)-value (\(SAOD\)) & \textless{} 0.001 & \textless{} 0.001 &
\textless{} 0.001 & 0.3196 & 0.4972 \\
\(RONI\) (lag 1) & 0.7504 & 0.7475 & 0.7478 & 0.6371 & 0.6316 \\
SE (\(RONI\), lag 1) & 0.1458 & 0.1452 & 0.1409 & 0.1637 & 0.1648 \\
\(p\)-value (\(RONI\), lag 1) & \textless{} 0.001 & \textless{} 0.001 &
\textless{} 0.001 & \textless{} 0.001 & \textless{} 0.001 \\
\(R\)-squared & 0.7531 & 0.7520 & 0.7531 & 0.6938 & 0.6921 \\
\end{longtable}

The natural-variability-adjusted growth series is obtained by
subtracting the fitted natural-variability component from the observed
growth, and the resulting airborne-fraction series is shown in
Figure~\ref{fig-data} center left. The three Pinatubo years in the
adjusted series are conspicuously high relative to the unadjusted
series. Smoothing them to the full-sample mean or to the non-Pinatubo
mean produces a more regular series that is less sensitive to the
leverage of a single episode.

\subsection{A rising airborne fraction across
estimators}\label{a-rising-airborne-fraction-across-estimators}

The main results are shown in Table~\ref{tbl-ladder}, estimating the
\(AF\) trend in the full sample \(1959\)--\(2024\). To isolate the role
of the \(1991\)--\(1993\) Mount Pinatubo eruption in the volcanic
adjustment, the table reports four growth series: the unadjusted series
(\(AF^{RAW}\)); the natural-variability-adjusted series with \(SAOD\) as
observed (\(AF^{ADJ}\), \(SAOD\)); and two adjusted series in which the
three Pinatubo years are smoothed by replacing their \(SAOD\) values
with the series mean, computed either over the full sample
(\(AF^{ADJ}\), \(SAOD^*\), full) or over the non-Pinatubo years
(\(AF^{ADJ}\), \(SAOD^*\), excl.). Each growth series is estimated by
all four estimators: \(OLS\), \(WLS\) and \(GLS\) on the collapsed
all-LULC mean series, and the \(MEM\) on the full \(69\)-series panel
(Methods).

\begin{longtable}[]{@{}
  >{\raggedright\arraybackslash}p{(\linewidth - 8\tabcolsep) * \real{0.4333}}
  >{\raggedleft\arraybackslash}p{(\linewidth - 8\tabcolsep) * \real{0.1000}}
  >{\raggedleft\arraybackslash}p{(\linewidth - 8\tabcolsep) * \real{0.1000}}
  >{\raggedleft\arraybackslash}p{(\linewidth - 8\tabcolsep) * \real{0.1667}}
  >{\raggedleft\arraybackslash}p{(\linewidth - 8\tabcolsep) * \real{0.2000}}@{}}
\caption{\(AF\) trend slope for the full sample (\(1959\)--\(2024\))
under the preferred natural-variability specification: \(RONI\) (lagged
one year) + \(SAOD\). Columns show the unadjusted growth series
(\(AF^{RAW}\)), the adjusted series with \(SAOD\) as observed
(\(AF^{ADJ}\), \(SAOD\)), and the adjusted series with the
\(1991\)--\(1993\) Pinatubo \(SAOD\) values replaced by the full-sample
mean (\(SAOD^*\), full) or the non-Pinatubo mean (\(SAOD^*\), excl.).
\(OLS\), \(WLS\) and \(GLS\) are estimated on the single all-LULC mean
\(AF\) series with \(HAC\) (Bartlett--Andrews) standard errors
\autocite{Andrews1991}; \(WLS\) and \(GLS\) weight each year by the
inverse delta-method variance of \(AF_t\), which propagates the
cross-series dispersion of the \(LULC\) denominator together with the
\(\pm 5\%\) fossil-emission uncertainty. The \(MEM\) uses the full
\(69\)-series panel with model-based standard errors. Intercepts are
reported in the Supplementary
Information.}\label{tbl-ladder}\tabularnewline
\toprule\noalign{}
\begin{minipage}[b]{\linewidth}\raggedright
\(AF\) trend slope (per year)
\end{minipage} & \begin{minipage}[b]{\linewidth}\raggedleft
\(AF^{RAW}\)
\end{minipage} & \begin{minipage}[b]{\linewidth}\raggedleft
\(AF^{ADJ}\) (SAOD)
\end{minipage} & \begin{minipage}[b]{\linewidth}\raggedleft
\(AF^{ADJ}\) (SAOD*, full)
\end{minipage} & \begin{minipage}[b]{\linewidth}\raggedleft
\(AF^{ADJ}\) (SAOD*, excl.)
\end{minipage} \\
\midrule\noalign{}
\endfirsthead
\toprule\noalign{}
\begin{minipage}[b]{\linewidth}\raggedright
\(AF\) trend slope (per year)
\end{minipage} & \begin{minipage}[b]{\linewidth}\raggedleft
\(AF^{RAW}\)
\end{minipage} & \begin{minipage}[b]{\linewidth}\raggedleft
\(AF^{ADJ}\) (SAOD)
\end{minipage} & \begin{minipage}[b]{\linewidth}\raggedleft
\(AF^{ADJ}\) (SAOD*, full)
\end{minipage} & \begin{minipage}[b]{\linewidth}\raggedleft
\(AF^{ADJ}\) (SAOD*, excl.)
\end{minipage} \\
\midrule\noalign{}
\endhead
\bottomrule\noalign{}
\endlastfoot
\textbf{\(OLS\)}: unweighted & & & & \\
Slope & 0.00117 & 0.00132 & 0.00164 & 0.00174 \\
SE (Slope) & 0.00064 & 0.00065 & 0.00053 & 0.00053 \\
\(p\)-value (Slope) & 0.069 & 0.041 & 0.002 & 0.001 \\
\textbf{\(WLS\)}: \(LULC\)-dispersion weighted & & & & \\
Slope & 0.00204 & 0.00199 & 0.00227 & 0.00238 \\
SE (Slope) & 0.00058 & 0.00054 & 0.00055 & 0.00058 \\
\(p\)-value (Slope) & \textless{} 0.001 & \textless{} 0.001 &
\textless{} 0.001 & \textless{} 0.001 \\
\textbf{\(GLS\)}: as \(WLS\), with \(AR(1)\) errors & & & & \\
Slope & 0.00221 & 0.00197 & 0.00237 & 0.00249 \\
SE (Slope) & 0.00062 & 0.00055 & 0.00059 & 0.00060 \\
\(p\)-value (Slope) & \textless{} 0.001 & \textless{} 0.001 &
\textless{} 0.001 & \textless{} 0.001 \\
\textbf{\(MEM\)}: 69-series panel, \(AR(1)\) errors & & & & \\
Slope & 0.00112 & 0.00120 & 0.00159 & 0.00170 \\
SE (Slope) & 0.00011 & 0.00008 & 0.00010 & 0.00010 \\
\(p\)-value (Slope) & \textless{} 0.001 & \textless{} 0.001 &
\textless{} 0.001 & \textless{} 0.001 \\
\end{longtable}

Our primary estimates come from the dispersion-weighted single-series
estimators, \(WLS\) and \(GLS\) (Table~\ref{tbl-ladder}), where the
latter allows for first-order autocorrelation in the residuals. Their
slopes are positive and statistically significant in every growth
specification, giving robust evidence of an increasing airborne fraction
that does not hinge on how the Pinatubo years are treated. Under the
preferred smoothed specifications the trend is about
\(0.0023\)--\(0.0024\) per year for \(WLS\) and \(0.0024\)--\(0.0025\)
for \(GLS\) (\(p<0.001\)); the unadjusted series gives a very similar
\(0.0020\) and \(0.0022\) (\(p<0.001\)); and even the specification that
leaves the eruption untreated in the continuous \(SAOD\) regressor
attenuates the slope only to about \(0.0020\) per year, where it remains
significant at the \(0.1\%\) level (\(p\approx 0.0002\)--\(0.0003\) for
both \(WLS\) and \(GLS\)). Because the weights are the inverse of a
delta-method variance that combines the cross-series \(LULC\) dispersion
with the reported \(\pm5\%\) fossil-emission uncertainty, they
downweight the poorly-measured early decades and anchor the fit to the
more reliable recent years, so the weighted estimates imply a somewhat
steeper rise, from about \(0.33\) in \(1959\) to about \(0.48\) in
\(2024\), than the unweighted and panel estimators, which place the
\(1959\) level nearer \(0.40\); all estimators agree that the airborne
fraction approaches \(0.47\)--\(0.50\) by \(2024\), broadly comparable
with the roughly \(0.44\) level reported in the literature
\autocite{Raupach2007,Knorr2009,Gloor2010CarbonFeedbackAF,bennedsenRegressionbasedApproachCO22024,bennedsenEvidenceTrendCO22023,veravaldes2025robustestimationco2}
while identifying a clear positive long-run trend.

The unweighted \(OLS\) estimates on the collapsed \(LULC\) mean tell the
same story with less power. The \(OLS\) slope is positive in every
growth specification and significant at the \(5\%\) level in all but
one: it reaches \(p=0.002\) and \(p=0.001\) once Pinatubo is smoothed,
\(p=0.04\) for the untreated-\(SAOD\) series, and softens to
non-significance only for the unadjusted series, \(p=0.07\). The
contrast with earlier work is instructive. Under the previous-generation
volcanic reconstruction, leaving the Pinatubo eruption untreated in the
continuous volcanic regressor collapsed the estimated \(OLS\) trend to
non-significance, the specification most vulnerable to the leverage of a
single episode, which helps explain why earlier studies have reported
weak or inconclusive evidence of an increasing \(AF\) trend
\autocite{Canadell2007,Raupach2007,Knorr2009,Ballantyne2012,LeQuere2009,bennedsenEvidenceTrendCO22023,bennedsenRegressionbasedApproachCO22024,bennettQuantificationAirborneFraction2024}.
With the updated \(CMIP7\) stratospheric-aerosol forcing, whose Pinatubo
peak is more moderate relative to the record, that fragility largely
disappears and the positive trend is recovered even without smoothing.
Nevertheless, the residual power loss of unweighted \(OLS\) is not
remedied by simply averaging more \(LULC\) series: simulation evidence
(Supplementary Information) shows that single-series and mean \(OLS\)
power is approximately constant in the number of \(LULC\) measurements
because the binding noise is common across the highly correlated series,
whereas the delta-method \(WLS\) and \(GLS\) estimators convert the
cross-series dispersion into reliability weights and gain power with the
number of series while holding valid coverage.

The \(MEM\), fitted to the full \(69\)-series panel, corroborates this
picture. Its fixed slope is positive and significant in every
specification and of comparable magnitude to the single-series
estimates, about \(0.0016\) per year under the smoothed specifications,
attenuating only to about \(0.0012\) for the untreated-\(SAOD\) series.
Through its random slopes, it confirms that the upward trend holds
within individual \(LULC\) definitions rather than only in their mean.
Its model-based standard errors are far smaller than those of the
weighted single-series estimators, but this apparent precision is not
additional evidence: because the \(69\) \(AF\) series share the
atmospheric-growth numerator and are \(\approx 98\%\) correlated, the
panel carries far less independent information than its \(69\) series
suggest, and the \(MEM\) standard errors correspondingly overstate it.
In simulation, its confidence intervals undercover severely as the panel
grows (Supplementary Information). We therefore treat the \(MEM\) as a
corroborating diagnostic and base inference on the weighted
single-series estimators.

\begin{figure}

\begin{minipage}[t]{0.50\linewidth}

\centering{

\pandocbounded{\includegraphics[keepaspectratio]{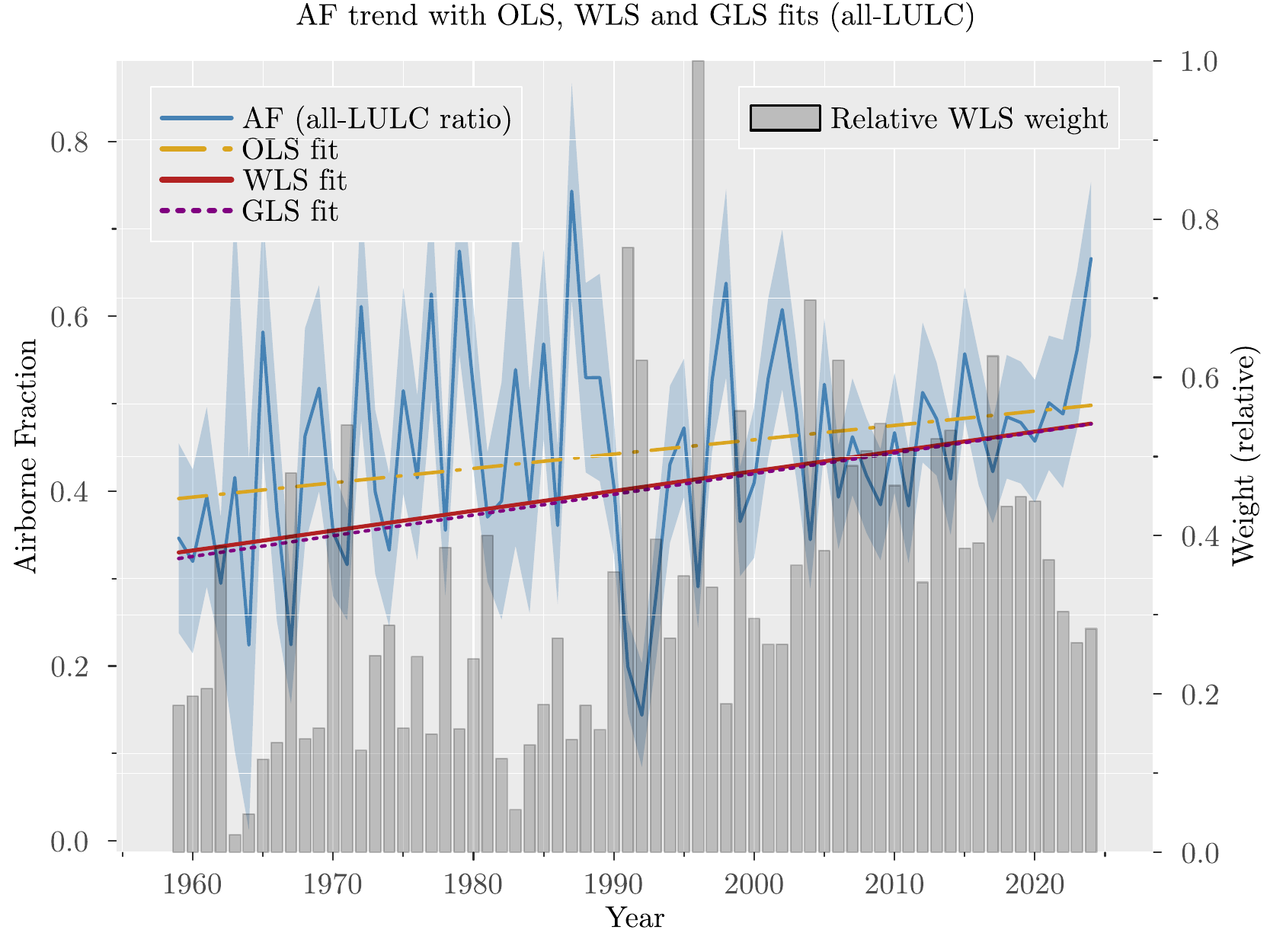}}

}

\subcaption{\label{fig-WLS-full-sample}Full sample, \(1959\)--\(2024\).}

\end{minipage}%
\begin{minipage}[t]{0.50\linewidth}

\centering{

\pandocbounded{\includegraphics[keepaspectratio]{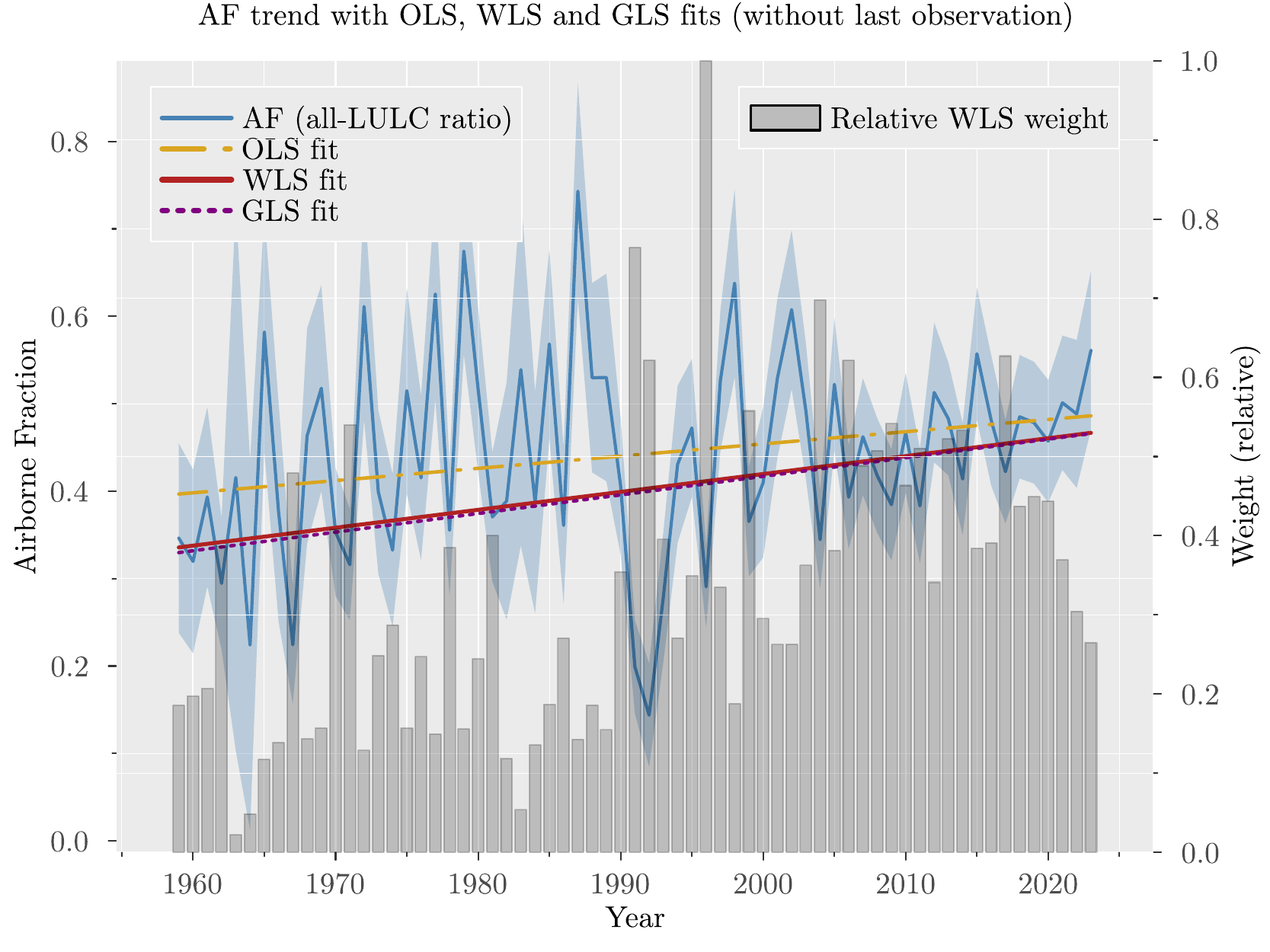}}

}

\subcaption{\label{fig-WLS-without-last}Final year (\(2024\)) excluded.}

\end{minipage}%

\caption{\label{fig-WLS-all-adjusted}Headline estimate and its endpoint
robustness. The \(LULC\) mean airborne fraction is shown with the
\(OLS\), \(WLS\), and \(GLS\) fitted trends and a delta-method 95\%
band; grey bars give the relative \(WLS\) weights. Preferred
natural-variability specification: contemporaneous \(SAOD\) and \(RONI\)
lagged one year, Pinatubo \(1991\)--\(93\) smoothed to the full-series
mean.}

\end{figure}%

Figure~\ref{fig-WLS-all-adjusted} presents the \(OLS\), \(WLS\), and
\(GLS\) fitted \(AF\) trends. The delta-method \(95\%\) confidence bands
are shown in light blue. The dispersion weights, grey bars, are broadly
smallest in the earlier years, where the \(LULC\) panel is most
dispersed and the annual \(AF\) estimate least reliable, so the weighted
fits are anchored by the better-measured recent decades. The conspicuous
exceptions are years such as \(1991\) and \(1996\), whose anomalously
low growth enters the delta-method variance through the numerator and
produces a large weight. Given that both sit near the sample midpoint,
however, they carry low leverage on the slope and shift only the
intercept. Both the \(WLS\) and \(GLS\) slopes are positive and
significant (Table~\ref{tbl-ladder}).

\begin{figure}

\begin{minipage}[t]{0.50\linewidth}

\centering{

\pandocbounded{\includegraphics[keepaspectratio]{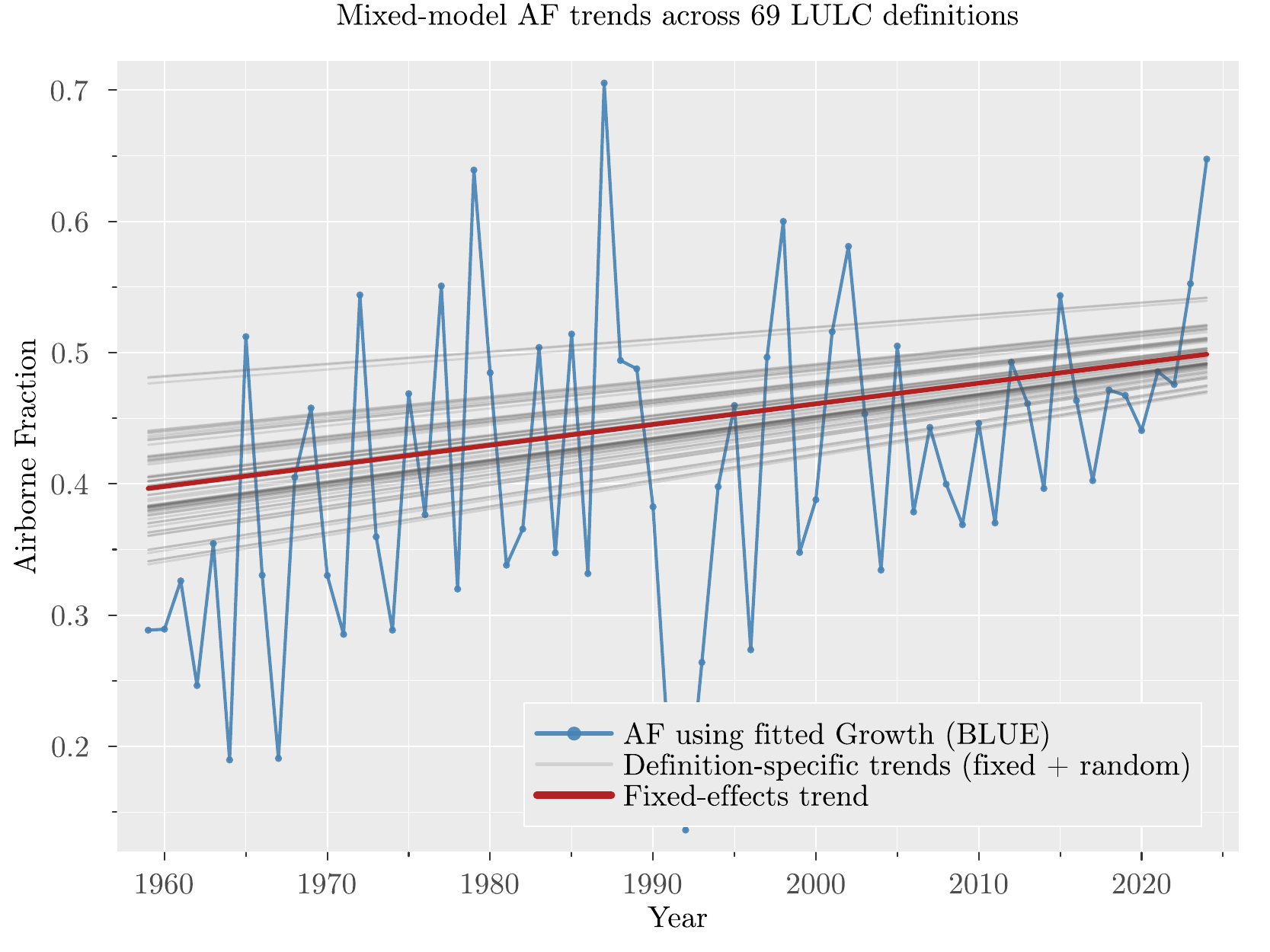}}

}

\subcaption{\label{fig-mixed-effects-trend}\(AF\) series with \(MEM\)
trend (natural-variability adjusted, full sample, \(1959\)--\(2024\)).}

\end{minipage}%
\begin{minipage}[t]{0.50\linewidth}

\centering{

\pandocbounded{\includegraphics[keepaspectratio]{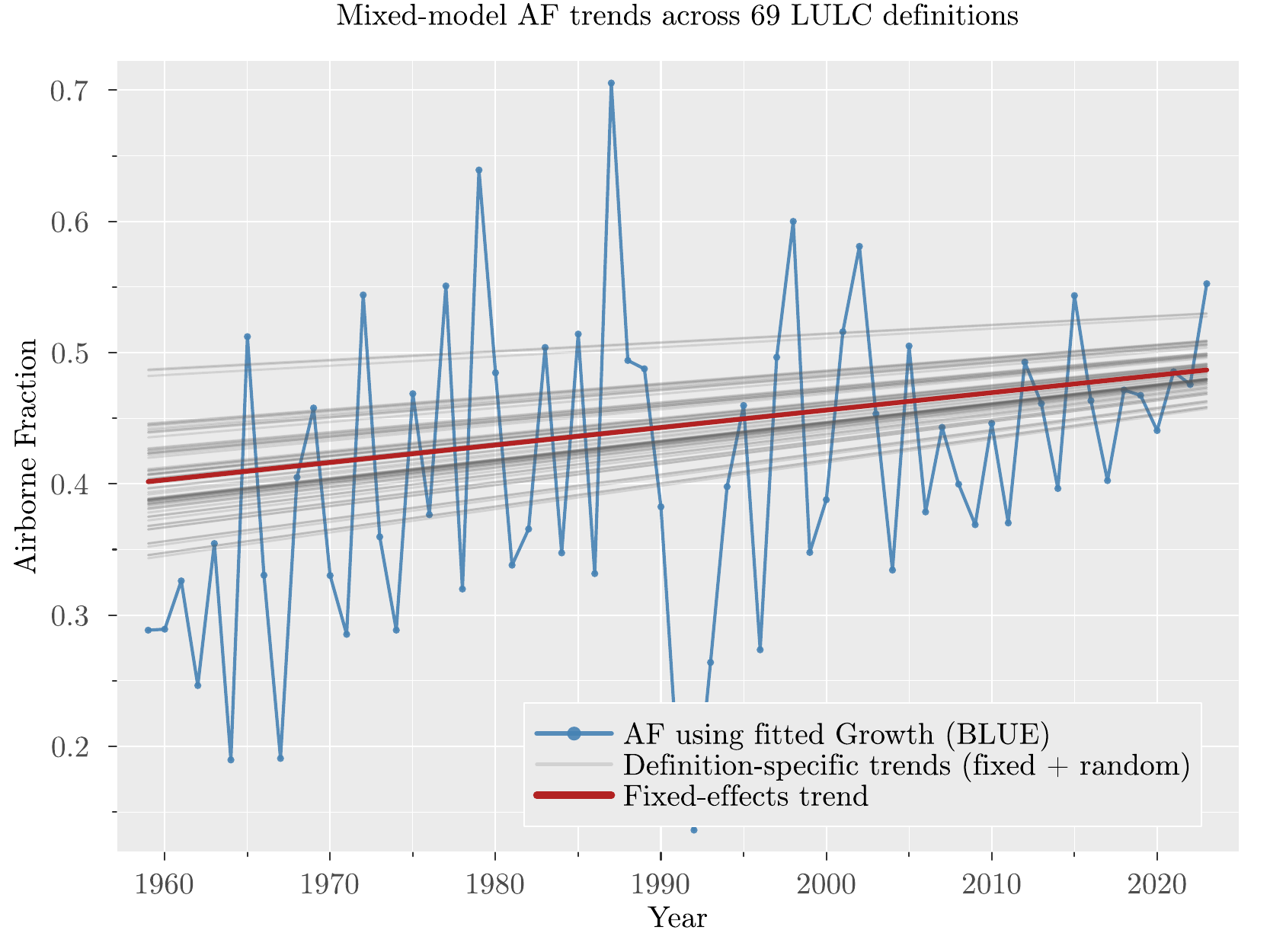}}

}

\subcaption{\label{fig-mixed-effects-trend-2023}\(AF\) series with
\(MEM\) trend (natural-variability adjusted, sample ending in
\(2023\)).}

\end{minipage}%

\caption{\label{fig-mixed-effects}Corroborating panel view: the \(AF\)
series and the pooled \(MEM\) trend across all \(69\) LULC measurement
series, for the full sample (left) and the sample ending in \(2023\)
(right). The trend is positive within individual definitions,
corroborating the weighted single-series headline estimate.}

\end{figure}

Figure~\ref{fig-mixed-effects-trend} shows the \(AF\) series for each
\(LULC\) measurement together with the pooled \(MEM\) trend for the
natural-variability-adjusted growth series over the full sample; one
representative series (\(BLUE\)) is highlighted for visual clarity,
while the trend is estimated on the full \(69\)-series ensemble. The
pooled trend and its series-specific slopes make the corroboration
visual: the positive trend of Figure~\ref{fig-WLS-all-adjusted} holds
within individual \(LULC\) definitions and not only in their mean.

\subsection{Endpoint and model specification
robustness}\label{endpoint-and-model-specification-robustness}

\(AF\) shows a large value in the final year of the sample (\(2024\)),
so we re-estimate the specifications of Table~\ref{tbl-ladder} on the
sample ending in \(2023\) to check whether the result is driven by that
observation (Figure~\ref{fig-WLS-without-last} and
Figure~\ref{fig-mixed-effects-trend-2023}). The pattern is unchanged:
the \(WLS\), \(GLS\) and \(MEM\) slopes remain positive and significant
at the \(0.1\%\) level in all four growth series. The slope estimates
are uniformly smaller than in the full sample, consistent with a
contribution from the large \(2024\) value, but the direction, the
Pinatubo-driven attenuation, and the inference are all preserved. The
\(OLS\) slope on the collapsed series remains significant only for the
smoothed specifications; the untreated-\(SAOD\) series, significant in
the full sample, now falls just short of the \(5\%\) level
(\(p\approx0.06\)). This result could explain the non-significance
reported in previous studies based on \(OLS\) and the untreated volcanic
index, which are more sensitive to the leverage of a single episode and
to the endpoint.

\begin{figure}

\centering{

\pandocbounded{\includegraphics[keepaspectratio]{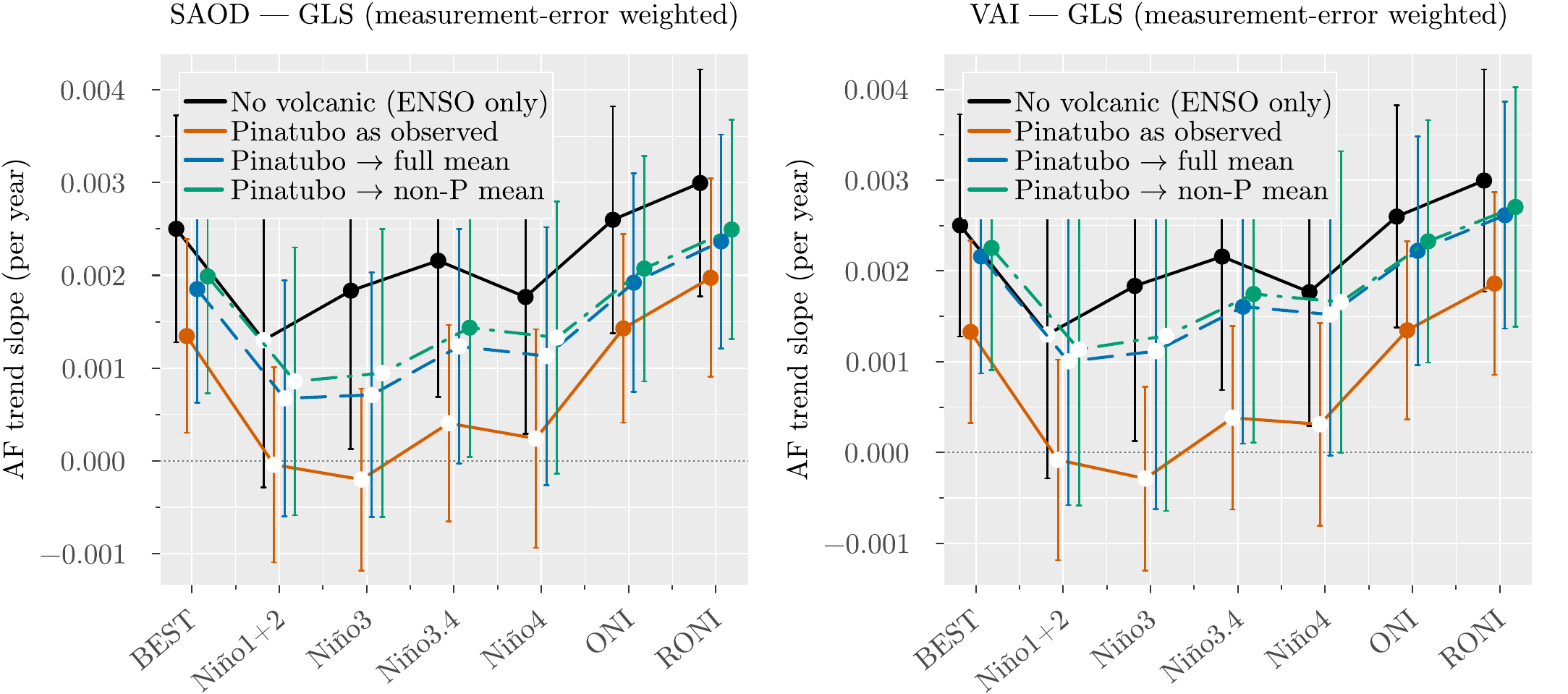}}

}

\caption{\label{fig-pinatubo-grid-gls}\(GLS\)-estimated \(AF\) trend
slopes across all natural-variability specifications, with delta-method
measurement-error weights. Each panel fixes the volcanic index (\(VAI\),
\(SAOD\)); the x-axis lists the seven \(ENSO\) indices (lagged one year)
and the four lines fix the volcanic treatment. Vertical bars are
\(95\%\) confidence intervals (\(\pm 1.96\) \(HAC\) standard errors),
dodged horizontally within each \(ENSO\) tick so they read as nudges
around each estimate; filled markers are significant at the \(5\%\)
level. Smoothing Pinatubo (blue, green) yields positive slopes that
reach significance for \(RONI\), \(ONI\), and \(BEST\) and positive
point estimates for the other indices. Leaving Pinatubo untreated
(orange) attenuates the slopes toward zero for the untreated Niño
indices, whose confidence intervals then include zero.}

\end{figure}

We next assess how sensitive the trend is to the natural-variability
specification by estimating all \(49\) combinations of the seven
\(ENSO\) indices (the indices of Figure~\ref{fig-data} top right, lagged
one year) with seven volcanic settings: no volcanic index, and the
\(VAI\) and \(SAOD\) each entered three ways (as observed, Pinatubo
smoothed to the full-sample mean, and Pinatubo smoothed to the
non-Pinatubo mean). Our primary robustness statement comes from the
measurement-error-weighted \(GLS\) estimator, whose delta-method
standard errors propagate the substantial first-stage uncertainty in the
volcanic coefficient into the \(AF\) trend.
Figure~\ref{fig-pinatubo-grid-gls} summarises the \(GLS\) slopes across
the grid. Once Pinatubo is smoothed, the \(GLS\) slope is positive in
every specification and significant at the \(5\%\) level for the
\(RONI\), \(ONI\), and \(BEST\) indices, with the largest and most
robust value given by \(RONI\), our preferred natural-variability
specification for the reasons set out above. The point estimate remains
positive for the four Niño sea-surface-temperature indices as well; they
are, however, estimated less precisely and mostly fall short of
significance, with only Niño 3.4 reaching the \(5\%\) level under the
non-Pinatubo mean. This is consistent with these raw
sea-surface-temperature boxes being noisier proxies of the latent
\(ENSO\) signal than the filtered \(ONI\), \(RONI\), and \(BEST\)
indices (Methods). Leaving Pinatubo untreated attenuates the slopes and,
for those untreated Niño indices, pulls the point estimate slightly
below zero; but because the \(GLS\) standard errors propagate the
first-stage volcanic uncertainty, these confidence intervals include
zero. The \(GLS\) slopes are somewhat larger than the corresponding
\(MEM\) slopes (Table~\ref{tbl-ladder}), consistent with the \(GLS\)
weighting down the earlier years that carry the greatest denominator
uncertainty.

The headline result does not depend on the expanded \(69\)-series
ensemble mean. Rebuilding the airborne-fraction series on the
conventional Global Carbon Budget denominator, the mean of the three
bookkeeping models (\(BLUE\), \(OSCAR\), \(LUCE\)), while keeping the
full-panel measurement-error weights leaves the \(WLS\) and \(GLS\)
trends positive and significant in every growth specification (slopes
\(\approx 0.0024\)--\(0.0028\) per year, all \(p<0.001\); Supplementary
Information). The positive trend is therefore a property of the data and
the weighting, not of the particular \(LULC\) ensemble mean used to
locate the point estimate.

Finally, each of the \(22\) process-based land models is augmented by
all three peat products (Methods) and so enters the \(69\)-series panel
three times, but this threefold representation does not drive the
result. The three peat estimates are small (\(\approx 0.2\) GtC
yr\(^{-1}\)) and mutually similar, so no peat variant is statistically
preferred, and re-estimating the full estimator ladder on reduced
\(25\)-series panels, each augmenting the \(22\) process models by only
one peat product, leaves the \(AF\) trend positive and significant
across \(OLS\), \(WLS\), \(GLS\), and \(MEM\) for every peat choice and
both samples, with the full-\(69\) estimate essentially
indistinguishable from the three (Supplementary Information). The trend
reflects the ensemble mean and its dispersion-based weighting, not the
repeated use of the process-model series.

\section{Discussion}\label{discussion}

Using a framework that controls for natural variability in the
atmospheric growth series and incorporates a broad \(LULC\)-definition
ensemble from Global Carbon Budget \(2025\), we find robust evidence
that \(AF\) increased from \(1959\) to \(2024\). Methodologically,
incorporating multi-source uncertainty materially changes inference
relative to plain \(OLS\). Endpoint tests show that this conclusion is
not driven by the large value in \(2024\).

A central methodological finding is that the divergence between studies
over the \(AF\) trend can be traced to a single event: the \(1991\)
Mount Pinatubo eruption. When atmospheric growth is filtered for natural
variability using a continuous volcanic aerosol index, three
high-leverage years dominate the volcanic coefficient. Leaving them
untreated attenuates the estimated trend for El Niño indices whose
sea-surface temperatures are contaminated by post-eruption Pacific
cooling, whereas smoothing those years or using the \(RONI\) index
restores a positive and significant trend. The apparent conflict in the
literature is therefore not about the underlying trend but about how a
single eruption is handled in the natural-variability adjustment and the
information loss of unweighted \(OLS\) relative to the weighted
single-series estimators. The attenuated slopes obtained under untreated
Pinatubo are small and, once the first-stage uncertainty is propagated
by \(WLS\) and \(GLS\), statistically indistinguishable from zero; they
arise only where post-eruption Pacific cooling contaminates the \(ENSO\)
index. They are thus a leverage artifact of the adjustment: a failure to
identify the trend rather than a finding against it.

Our preferred smoothed specifications imply that \(AF\) rose to about
\(0.47\)--\(0.50\) by \(2024\) (from roughly \(0.32\)--\(0.40\) around
\(1960\) across estimators), compared with the commonly reported value
of around \(0.44\). For a given emissions pathway, a higher \(AF\) would
be associated with faster atmospheric \(CO_2\) growth than under a
constant-\(AF\) baseline, potentially reducing the remaining carbon
budget for a given temperature target
\autocite{Canadell2007,Raupach2007,Friedlingstein2025}. IPCC AR6 carbon
budget assessments, built on Earth System Models, implicitly assume a
rising \(AF\) due to carbon-cycle feedbacks \autocite{IPCC_AR6_WG1}, but
observational confirmation of this trend has been elusive. Our results
provide that confirmation. A rising airborne fraction also sits
alongside the independently documented acceleration of Earth's energy
imbalance and of ocean and surface warming over recent decades
\autocite{Miniere2023,StortoYang2024,rahmstorfGlobalWarmingHas2025,WMO_SGC_2025,VeraValdes2026acceleration,bugajski2026spatialemergenceaccelerationglobal}:
a larger share of emissions remaining airborne means faster \(CO_2\)
accumulation and stronger radiative forcing, so the two lines of
evidence are mutually reinforcing.

Applied to the IPCC AR6 remaining budget of \(300\)--\(400\) \(GtCO_2\)
\autocite{IPCC_AR6_WG1}, the observed rise in \(AF\) to about
\(0.47\)--\(0.50\), some \(0.04\)--\(0.06\) above the historical
observational baseline of \(0.44\)
\autocite{Canadell2007,Raupach2007,Knorr2009}, implies an effective
reduction of approximately \(23\)--\(48\) \(GtCO_2\), up to roughly one
year of current global fossil emissions (see Supplementary Information
for the derivation and caveats). This conclusion holds across
natural-variability specifications provided the volcanic adjustment does
not let the Pinatubo eruption dominate: across the grid of volcanic and
\(ENSO\) indices, the estimated \(AF\) trend is positive once the
eruption is smoothed or a volcanically robust \(ENSO\) index is used,
and the weighted single-series (\(WLS\)/\(GLS\)) and \(MEM\) estimators
agree on a positive and significant trend under our preferred
specification.

\section{Methods}\label{methods}

\subsection{Data}\label{data}

We use annual Global Carbon Budget \(2025\) data, with atmospheric
growth \(G_t\) from NOAA/ESRL global concentration trends
\autocite{Lan2025}, fossil emissions excluding carbonation \(FF_t\) from
the Global Carbon Project fossil dataset \autocite{Friedlingstein2025},
and a panel of \(69\) \(LULC\) measurements per year: \(BLUE\)
\autocite{Hansis2015}, \(OSCAR\) \autocite{Gasser2020}, \(LUCE\)
\autocite{Qin2024}, and peat-augmented
\autocite{Conchedda2020,Mueller2021,Qiu2021} process-based land-model
combinations drawn from the model ensemble
\autocite{Haverd2018,Melton2020,Lawrence2019,Fisher2015,Tian2015,Ma2022,Yang2023,Needham2025,Felzer2018,Xia2024,Yue2024,Shu2020,Reick2021,Poulter2011,Smith2014,Schaphoff2018,Lienert2018,Vuichard2019,Walker2017,Kato2013,Ito2019}.
Volcanic activity is measured primarily by the global-mean stratospheric
aerosol optical depth (\(SAOD\)) at \(550\) nm and, as a secondary index
in the sensitivity grid, by the Volcanic Aerosol Index (\(VAI\))
\autocite{VAI2003}. The \(SAOD\) series is built end-to-end from a
single, consistent satellite-era product (\(SAOD\) construction, below).
\(ENSO\) variability is measured by seven indices from the NOAA Physical
Sciences Laboratory \autocite{NOAA_PSL_ENSO}: the Niño 1+2, Niño 3, Niño
3.4, and Niño 4 sea-surface-temperature indices, the Oceanic Niño Index
(\(ONI\)), the relative Oceanic Niño Index (\(RONI\))
\autocite{LHeureux2024RONI,Huang_2017_ERSSTv5_JCLI}, and the Bivariate
\(ENSO\) Timeseries (\(BEST\)) \autocite{Smith2000}.

\subsubsection*{Land-use and land-cover change (LULC)
data}\label{land-use-and-land-cover-change-lulc-data}
\addcontentsline{toc}{subsubsection}{Land-use and land-cover change
(LULC) data}

The \(69\) \(LULC\) measurements are built from the GCB \(2025\)
dataset. We first extract the three bookkeeping series (\(BLUE\),
\(OSCAR\), \(LUCE\)). Then, for each process-based land-model series in
the GCB ensemble that does not already include peat emissions, we add
one peat component to make it comparable to the bookkeeping models,
which include peat by construction: for each of the \(22\) process-based
series we form three peat-augmented variants by adding FAO\_peat,
LPX\_Bern\_peat, and ORCHIDEE\_peat
\autocite{Conchedda2020,Mueller2021,Qiu2021} respectively. This produces
\(66\) derived series, which together with \(BLUE\)/\(OSCAR\)/\(LUCE\)
give the panel of \(69\) yearly \(LULC\) measurements.

\subsubsection*{Volcanic indices data}\label{volcanic-indices-data}
\addcontentsline{toc}{subsubsection}{Volcanic indices data}

The stratospheric aerosol optical depth series is rebuilt end-to-end.
For \(1959\)--\(2023\) we use the CMIP7 v2.2.1 stratospheric aerosol
forcing \autocite{Aubry2025CMIP7,Aubry2025CMIP7data}, the University of
Exeter \texttt{input4MIPs} dataset prepared for the seventh Coupled
Model Intercomparison Project, which provides the zonal-mean
stratospheric aerosol optical depth at \(550\) nm under one methodology
spanning the Agung (\(1963\)), El Chichón (\(1982\)), Pinatubo
(\(1991\)), and Hunga Tonga (\(2022\)) eruptions. We collapse the
monthly zonal-mean field to an annual global mean by area
(cosine-latitude) weighting. Because the CMIP7 release ends in \(2023\),
the single remaining year (\(2024\)) is appended from GloSSAC v2.24
\autocite{Kovilakam2020GloSSAC,Kovilakam2026GloSSACdata}, the NASA
Global Space-based Stratospheric Aerosol Climatology, processed
identically to a \(550\) nm global mean and rescaled onto the CMIP7
level using their \(2013\)--\(2023\) overlap; the two products are the
same \(550\) nm satellite global mean and agree to within a scale factor
of about \(0.9\), so this is a seamless one-year extension rather than a
splice. The \(2024\) value still carries the decaying Hunga Tonga
aerosol and replaces the zero previously assumed for that year. A
constant rescaling of the \(SAOD\) regressor is absorbed by its
first-stage coefficient and leaves the adjusted growth series unchanged,
so the choice of absolute level does not affect inference.

\(VAI\) is a secondary volcanic index used in the sensitivity grid.
\(VAI\) has not been updated since \(2021\) and we pad it with zeros for
\(2022\)--\(2024\), which is consistent with the low \(SAOD\) values in
those years. The \(VAI\) series is used only in the sensitivity grid.

\subsection{Natural-variability
adjustment}\label{natural-variability-adjustment}

Natural factors such as volcanic eruptions and El Niño events can cause
year-to-year fluctuations in atmospheric \(CO_2\) growth that are not
directly related to anthropogenic emissions. To account for this natural
variability, the literature considers a specification that includes a
control for \(ENSO\) and volcanic activity
\autocite{LeQuere2009,bennedsenRegressionbasedApproachCO22024,bennettQuantificationAirborneFraction2024,veravaldes2025robustestimationco2}.
The preferred natural-variability-adjusted growth regression
specification is as follows:
\begin{equation}\protect\phantomsection\label{eq-growth-reg}{
\hat{G}_t = \gamma_0 + \gamma_1 \hat{C}_t + \gamma_2 ENSO_{t-1} + \gamma_3 SAOD_t + \varepsilon_t,
}\end{equation}

where \(\hat{C}_t\) is the estimated total emissions (fossil fuel plus
\(LULC\)), \(ENSO_{t-1}\) is the one-year-lagged \(ENSO\) index
(\(RONI\) in the preferred specification), and \(SAOD_t\) is the
stratospheric aerosol optical depth for year \(t\). The error term
captures idiosyncratic noise.

Equation~\ref{eq-growth-reg} is typically estimated by ordinary least
squares (\(OLS\)), but \(OLS\) ignores the measurement uncertainty in
total anthropogenic emissions \(\hat{C}_t\). We instead estimate it by
weighted least squares (\(WLS\)), weighting by the inverse variance of
\(\hat{C}_t\), which combines two independent components: the
cross-model dispersion of the \(69\)-series \(LULC\) panel and the
Global Carbon Budget's reported \(\pm5\%\) (\(1\sigma\)) uncertainty on
fossil emissions, \[
\operatorname{\widehat{Var}}(\hat{C}_t) = \operatorname{\widehat{Var}}\bigl(LULC_t\bigr) + \operatorname{\widehat{Var}} \bigl(FF_t\bigr) .
\]

Given that the \(FF_t\) and \(LULC_t\) terms derive from independent
data (energy statistics versus bookkeeping and land models), their
covariance is assumed negligible. The trend estimates are insensitive to
the assumed magnitude of the fossil-emission uncertainty over the
plausible range \(0\)--\(7\%\) (Supplementary Information). Propagating
\(\hat{C}_t\) through the delta method (Supplementary Information), the
measurement-error contribution to the variance of growth in year \(t\)
is approximately
\(\gamma_1^2\,\operatorname{\widehat{Var}}(\hat{C}_t)\); adding the
idiosyncratic residual variance \(\sigma^2\) gives the total conditional
variance of the growth observation,

\[
\omega_t^2 = \sigma^2 + \gamma_1^2 \operatorname{\widehat{Var}}(\hat{C}_t).
\]

The \(WLS\) weights are the reciprocal of this variance, downweighting
years in which the \(LULC\) definitions disagree most about total
emissions. Under heteroskedasticity this is efficient and yields valid
inference, whereas \(OLS\), which assumes a constant annual variance, is
not. To control for residual autocorrelation we also estimate a
generalised least squares (\(GLS\)) specification with an autoregressive
error structure, using the same controls and weights as \(WLS\). Because
the \(69\) \(LULC\) series share input datasets and modelling
assumptions, their cross-series dispersion can be thought of as a lower
bound on the structural uncertainty in land-use emissions. Yet, this
affects only the relative weights. Since \(WLS\) and \(GLS\) are
invariant to the overall scale of the weights and inference rests on
\(HAC\) standard errors computed from the fitted residuals rather than
on the assumed dispersion level, the overall magnitude of the
denominator uncertainty is immaterial to both the point estimates and
their significance.

Our preferred specification uses \(RONI\), given that it removes a
tropical-mean sea-surface-temperature signal and is therefore robust to
the transient Pacific cooling that follows large volcanic eruptions (the
remaining indices are used in the sensitivity grid of
Figure~\ref{fig-pinatubo-grid-gls}). The \(ENSO\) index enters with a
one-year lag, which improves the fit relative to the contemporaneous
specification (Supplementary Information) and is consistent with the
documented delay between \(ENSO\) sea-surface-temperature anomalies and
their effect on the tropical land and ocean carbon flux
\autocite{Betts2016}. This lag is selected by first-stage fit: the
\(R^2\) rises from \(0.695\) at lag zero to \(0.753\) at lag one, and
lag one is either optimal or fit-improving for six of the seven \(ENSO\)
indices (Supplementary Information). The volcanic index fits best
contemporaneously and enters without a lag.

The \(1991\)--\(1993\) Mount Pinatubo eruption produced aerosol values
much larger than any other event in the sample (Figure~\ref{fig-data}
top left), so we assess how the natural-variability adjustment depends
on this single event. In addition to the raw and as-observed
specifications, we construct two smoothed volcanic series in which the
\(SAOD\) (or \(VAI\)) values for \(1991\), \(1992\), and \(1993\) are
replaced by the series mean, computed either over the full sample or
over the non-Pinatubo years. This mitigates the high-leverage Pinatubo
years without discarding any observation from the sample. Replacing
these three values with the series mean is a transparent, standard
device for limiting the leverage of an outlying episode, closely
analogous to including indicator (dummy) variables for
\(1991\)--\(1993\); it differs only in that it modifies the volcanic
regressor rather than adding columns, and so curbs the eruption's
influence on the volcanic coefficient while retaining every observation
for the estimation of the remaining terms.

The natural-variability-adjusted growth series is constructed by
subtracting the fitted values of the natural-variability controls from
the observed growth series,
\(\hat{G}_t^{ADJ} = G_t - \hat{\gamma}_2 \text{$ENSO$}_{t-1} - \hat{\gamma}_3 \text{SAOD}_t\),
where \(G_t\) is observed growth and the coefficients are taken from the
\(GLS\) fit. Only the \(ENSO\) and volcanic contributions are removed,
so the emissions-driven component and the residual are retained. The raw
and natural-variability-adjusted growth series are shown in
Figure~\ref{fig-data} centre left.

Besides the preferred specification, we also consider a sensitivity grid
of specifications that use all combinations of the two volcanic indices
(\(VAI\) and \(SAOD\)) and the seven \(ENSO\) indices (Niño 1+2, Niño 3,
Niño 3.4, Niño 4, \(ONI\), \(RONI\), and \(BEST\)). The results of this
grid are reported in Figure~\ref{fig-pinatubo-grid-gls}.

The seven \(ENSO\) indices differ in construction, and this difference
maps onto how much reliable evidence each provides for the trend. The
four Niño indices (Niño 1+2, Niño 3, Niño 3.4, Niño 4) are simple
averages of sea-surface-temperature anomalies over fixed
equatorial-Pacific boxes, taken relative to a fixed climatological base
and with no adjustment for the tropical-mean warming background. By
contrast, \(ONI\) is a three-month running mean of Niño 3.4 anomalies
with a periodically updated base period, \(BEST\) combines the Niño 3.4
signal with the atmospheric Southern Oscillation Index into a single
coupled ocean--atmosphere measure \autocite{Smith2000}, and \(RONI\)
further subtracts the contemporaneous tropical-mean
sea-surface-temperature anomaly so that the shared warming signal is
removed \autocite{LHeureux2024RONI}, making it our preferred index for
the natural-variability adjustment. The raw Niño boxes are therefore
noisier and more contaminated by local variability and by the evolving
mean state of the tropical Pacific, so they proxy the latent \(ENSO\)
signal with more effective measurement error. This could explain why the
\(AF\) trend is least precisely estimated for these indices
(Figure~\ref{fig-pinatubo-grid-gls}): the instability is not evidence
against the trend but a reflection of the weaker, noisier
natural-variability adjustment they afford, and it is most consequential
when the transient post-Pinatubo Pacific cooling projects onto the
sea-surface-temperature boxes and, left uncorrected, masquerades as an
\(ENSO\) signal.

\subsection{Dispersion-weighted airborne-fraction
trend}\label{dispersion-weighted-airborne-fraction-trend}

The empirical question is whether \(AF\) has been rising over the sample
period, which is equivalent to testing for a positive linear trend. For
each time \(t\), we obtain a point estimate \(AF_t\) and an uncertainty
proxy that reflects cross-measurement dispersion. We then estimate a
linear trend in \(AF\) over time, allowing for heteroskedasticity in the
annual variance of \(AF_t\):

\begin{equation}\protect\phantomsection\label{eq-af-trend}{
AF_t = \alpha + \beta t + \varepsilon_t, \quad \varepsilon_t \sim (0, \sigma_t^2), \quad \sigma_t^2 = \operatorname{Var}(AF_t).
}\end{equation}

Testing for a time trend is then a test of \(\beta=0\) versus
\(\beta\neq 0\) in Equation~\ref{eq-af-trend}.

Previous studies estimate the trend by \(OLS\) on the collapsed series
of \(AF\) estimates, or by univariate regressions on each \(LULC\)
measurement separately. Both approaches discard the cross-measurement
information in the denominator and yield fragile inference. In terms of
Equation~\ref{eq-af-trend}, this amounts to assuming that the variance
\(\sigma_t^2\) is constant across years; if it is not, \(OLS\) is
inefficient and its inference invalid.

The key design choice is how to use the disagreement among the multiple
\(LULC\) measurements without treating the highly correlated series as
independent. Collapsing the panel to a single series and estimating the
trend by \(OLS\) discards the cross-measurement information and yields
fragile, low-power inference (Table~\ref{tbl-ladder}). Pooling all
\(69\) series as if they were independent replicates instead overstates
the information they carry, because the series share the
atmospheric-growth numerator and are \(\approx 98\%\) correlated across
\(LULC\) definitions. Our preferred strategy takes a middle path: it
collapses the panel to a single annual airborne-fraction series but
propagates the cross-measurement dispersion of the denominator into the
trend estimate through a two-stage \(WLS\) procedure.

Given the adjusted growth series, we form a single annual
airborne-fraction series with the \(LULC\) mean denominator and estimate
its trend by \(WLS\), weighting each year by the inverse of a
delta-method variance. That variance propagates uncertainty from both
the numerator and the denominator, and for the ratio in
Equation~\ref{eq-af-def} it is approximately:

\begin{equation}\protect\phantomsection\label{eq-af-var}{
\widehat{\operatorname{Var}}(AF_t) = \left(\frac{1}{\hat{C}_t}\right)^2 \widehat{\operatorname{Var}}(\hat{G}_t) + \left(\frac{\hat{G}_t}{\hat{C}_t^2}\right)^2 \widehat{\operatorname{Var}}(\hat C_t),
}\end{equation}

where \(\hat{C}_t\) is the \(LULC\) mean denominator and
\(\widehat{\operatorname{Var}}(\hat{G}_t)\) is the sampling variance of
the natural-variability-adjusted growth,
\(z_t^\top\,\widehat{\operatorname{Cov}}(\hat\gamma)\,z_t\), with
\(z_t\) the first-stage design row restricted to the two
natural-variability columns that are actually subtracted and
\(\widehat{\operatorname{Cov}}(\hat\gamma)\) the coefficient covariance
of the \(GLS\) fit those coefficients come from (zero for the unadjusted
\(AF^{RAW}\) series, whose growth is observed rather than estimated;
Supplementary Information); the first term is the numerator contribution
and the second the denominator contribution. Weighting by the inverse of
this variance gives more weight to years with a better-measured
denominator, so the trend is anchored by the more reliable recent
decades. A \(GLS\) variant adds a Prais--Winsten autoregressive error
structure for residual autocorrelation \autocite{PraisWinsten1954}.
Throughout, inference uses heteroskedasticity- and
autocorrelation-consistent (\(HAC\)) standard errors
\autocite{NeweyWest1987,Andrews1991}.

\subsection{Mixed-effects cross-check}\label{mixed-effects-cross-check}

As a cross-check on the single-series design, we also fit a
mixed-effects model (\(MEM\)) to the full panel of \(69\) yearly \(AF\)
series \autocite{Henderson1953,Bolker2009}. For each LULC measurement
series we construct a series of \(AF\) estimates by year
(Equation~\ref{eq-af-def}), and the model estimates a common time trend
across all series while allowing for series-specific intercepts and
slopes. The model is estimated by restricted maximum likelihood, with an
\(AR(1)\) within-series error structure and model-based standard errors.
Given that the \(69\) series share the growth numerator and are
near-collinear, we read it as a corroborating diagnostic rather than the
primary inference (full specification in the Supplementary Information).

\section*{Acknowledgements}\label{acknowledgements}
\addcontentsline{toc}{section}{Acknowledgements}

This research received no specific grant from any funding agency in the
public, commercial, or not-for-profit sectors.

\section*{Author contributions}\label{author-contributions}
\addcontentsline{toc}{section}{Author contributions}

J.E.V.-V. conceived the study, performed the analysis, and wrote the
manuscript.

\section*{Conflict of Interest
Statement}\label{conflict-of-interest-statement}
\addcontentsline{toc}{section}{Conflict of Interest Statement}

The author declares no conflicts of interest.

\section*{Open Research}\label{open-research}
\addcontentsline{toc}{section}{Open Research}

\textbf{Data Availability Statement.} The data that support the findings
of this study are openly available. The Global Carbon Budget \(2025\)
data are available at \url{https://globalcarbonbudget.org/gcb-2025/};
the \(ENSO\) indices from the NOAA Physical Sciences Laboratory
(\url{https://psl.noaa.gov/enso/}); the CMIP7 v2.2.1 stratospheric
aerosol forcing used for the \(SAOD\) series from the input4MIPs archive
\autocite{Aubry2025CMIP7data} (University of Exeter;
\url{https://aims2.llnl.gov/search/input4MIPs/}), extended to \(2024\)
with GloSSAC v2.24 \autocite{Kovilakam2026GloSSACdata} from the NASA
Atmospheric Science Data Center
(\url{https://asdc.larc.nasa.gov/project/GloSSAC}); and the Volcanic
Aerosol Index from the NOAA National Centers for Environmental
Information
(\url{https://www.ncei.noaa.gov/access/paleo-search/study/5786}).

\textbf{Software Availability Statement.} All analysis code that
reproduces the results, figures, and tables in this study is openly
available in a GitHub repository at
\url{https://github.com/everval/Airborne-Fraction-Multiple-LULC-Measurements/}
under the DOI:
\href{https://doi.org/10.5281/zenodo.21543771}{10.5281/zenodo.21543771}.

\clearpage
\setcounter{figure}{0}\setcounter{table}{0}\setcounter{equation}{0}
\renewcommand{\thefigure}{S\arabic{figure}}
\renewcommand{\thetable}{S\arabic{table}}
\renewcommand{\theequation}{S\arabic{equation}}
\begin{center}
{\Large\textbf{Supplementary information}}
\end{center}
\bigskip

\section{Estimator ladder: intercepts and subsample
estimates}\label{estimator-ladder-intercepts-and-subsample-estimates}

This section reports the detail underlying the estimator ladder table of
the main text: the intercepts for all four estimators on the full
sample, and the complete ladder re-estimated on the subsample ending in
\(2023\). All estimates use the preferred \(RONI\) (lagged one year) +
\(SAOD\) natural-variability specification.

The intercepts (Table~\ref{tbl-ladder-intercepts}) reveal a systematic
pattern. The \(WLS\) and \(GLS\) intercepts sit below the \(OLS\) and
\(MEM\), around \(0.32\) against \(0.39\) for the smoothed
specifications, because the measurement-error weights are smallest in
the early years, where the cross-series dispersion of the \(LULC\)
denominator is largest. Those estimators therefore pin the fitted line
on the well-measured later years. The difference between the
\(WLS\)/\(GLS\) and \(MEM\) intercepts is well within one standard error
of the former, so the two are not in conflict; the trend slope, not the
level at the start of the sample, is what the weighted single-series
estimators identify sharply.

\begin{longtable}[]{@{}
  >{\raggedright\arraybackslash}p{(\linewidth - 8\tabcolsep) * \real{0.2593}}
  >{\raggedleft\arraybackslash}p{(\linewidth - 8\tabcolsep) * \real{0.1852}}
  >{\raggedleft\arraybackslash}p{(\linewidth - 8\tabcolsep) * \real{0.1852}}
  >{\raggedleft\arraybackslash}p{(\linewidth - 8\tabcolsep) * \real{0.1852}}
  >{\raggedleft\arraybackslash}p{(\linewidth - 8\tabcolsep) * \real{0.1852}}@{}}
\caption{Intercepts by estimator and growth specification, full sample
(\(1959\)--\(2024\)), for the preferred \(RONI\) (lagged one year) +
\(SAOD\) specification. Columns and estimators are as in the estimator
ladder table of the main text. Every intercept is significant at
\(p<0.001\).}\label{tbl-ladder-intercepts}\tabularnewline
\toprule\noalign{}
\begin{minipage}[b]{\linewidth}\raggedright
Intercept
\end{minipage} & \begin{minipage}[b]{\linewidth}\raggedleft
\(AF^{RAW}\)
\end{minipage} & \begin{minipage}[b]{\linewidth}\raggedleft
\(AF^{ADJ}\) (SAOD)
\end{minipage} & \begin{minipage}[b]{\linewidth}\raggedleft
\(AF^{ADJ}\) (SAOD*, full)
\end{minipage} & \begin{minipage}[b]{\linewidth}\raggedleft
\(AF^{ADJ}\) (SAOD*, excl.)
\end{minipage} \\
\midrule\noalign{}
\endfirsthead
\toprule\noalign{}
\begin{minipage}[b]{\linewidth}\raggedright
Intercept
\end{minipage} & \begin{minipage}[b]{\linewidth}\raggedleft
\(AF^{RAW}\)
\end{minipage} & \begin{minipage}[b]{\linewidth}\raggedleft
\(AF^{ADJ}\) (SAOD)
\end{minipage} & \begin{minipage}[b]{\linewidth}\raggedleft
\(AF^{ADJ}\) (SAOD*, full)
\end{minipage} & \begin{minipage}[b]{\linewidth}\raggedleft
\(AF^{ADJ}\) (SAOD*, excl.)
\end{minipage} \\
\midrule\noalign{}
\endhead
\bottomrule\noalign{}
\endlastfoot
\textbf{\(OLS\)} & & & & \\
Intercept & 0.4019 & 0.4285 & 0.3915 & 0.3838 \\
SE (Intercept) & 0.0283 & 0.0288 & 0.0240 & 0.0245 \\
\textbf{\(WLS\)} & & & & \\
Intercept & 0.3102 & 0.3696 & 0.3299 & 0.3180 \\
SE (Intercept) & 0.0339 & 0.0243 & 0.0312 & 0.0361 \\
\textbf{\(GLS\)} & & & & \\
Intercept & 0.2966 & 0.3706 & 0.3230 & 0.3096 \\
SE (Intercept) & 0.0347 & 0.0246 & 0.0321 & 0.0369 \\
\textbf{\(MEM\)} & & & & \\
Intercept & 0.4068 & 0.4353 & 0.3960 & 0.3881 \\
SE (Intercept) & 0.0057 & 0.0057 & 0.0056 & 0.0055 \\
\end{longtable}

Re-estimating the full ladder on the sample ending in \(2023\)
(Table~\ref{tbl-ladder-2023}) leaves every qualitative conclusion
intact. The slopes are uniformly smaller than in the full sample,
consistent with a contribution from the large \(2024\) value, but the
ordering across estimators, the attenuation under the untreated-\(SAOD\)
specification, and the inference are all preserved: \(OLS\) is
non-significant for both the unadjusted and the untreated-\(SAOD\)
series, \(WLS\) and \(GLS\) remain significant at \(p<0.001\) in every
column, and the \(MEM\) remains significant at \(p<0.001\) throughout.

\begin{longtable}[]{@{}
  >{\raggedright\arraybackslash}p{(\linewidth - 8\tabcolsep) * \real{0.2593}}
  >{\raggedleft\arraybackslash}p{(\linewidth - 8\tabcolsep) * \real{0.1852}}
  >{\raggedleft\arraybackslash}p{(\linewidth - 8\tabcolsep) * \real{0.1852}}
  >{\raggedleft\arraybackslash}p{(\linewidth - 8\tabcolsep) * \real{0.1852}}
  >{\raggedleft\arraybackslash}p{(\linewidth - 8\tabcolsep) * \real{0.1852}}@{}}
\caption{\(AF\) trend slope and intercept by estimator and growth
specification, subsample \(1959\)--\(2023\), for the preferred \(RONI\)
(lagged one year) + \(SAOD\) specification. Columns, estimators, and
standard-error conventions are as in the estimator ladder table of the
main text.}\label{tbl-ladder-2023}\tabularnewline
\toprule\noalign{}
\begin{minipage}[b]{\linewidth}\raggedright
AF trend slope (per year)
\end{minipage} & \begin{minipage}[b]{\linewidth}\raggedleft
\(AF^{RAW}\)
\end{minipage} & \begin{minipage}[b]{\linewidth}\raggedleft
\(AF^{ADJ}\) (SAOD)
\end{minipage} & \begin{minipage}[b]{\linewidth}\raggedleft
\(AF^{ADJ}\) (SAOD*, full)
\end{minipage} & \begin{minipage}[b]{\linewidth}\raggedleft
\(AF^{ADJ}\) (SAOD*, excl.)
\end{minipage} \\
\midrule\noalign{}
\endfirsthead
\toprule\noalign{}
\begin{minipage}[b]{\linewidth}\raggedright
AF trend slope (per year)
\end{minipage} & \begin{minipage}[b]{\linewidth}\raggedleft
\(AF^{RAW}\)
\end{minipage} & \begin{minipage}[b]{\linewidth}\raggedleft
\(AF^{ADJ}\) (SAOD)
\end{minipage} & \begin{minipage}[b]{\linewidth}\raggedleft
\(AF^{ADJ}\) (SAOD*, full)
\end{minipage} & \begin{minipage}[b]{\linewidth}\raggedleft
\(AF^{ADJ}\) (SAOD*, excl.)
\end{minipage} \\
\midrule\noalign{}
\endhead
\bottomrule\noalign{}
\endlastfoot
\textbf{\(OLS\)} & & & & \\
Slope & 0.00088 & 0.00109 & 0.00140 & 0.00150 \\
SE (Slope) & 0.00060 & 0.00058 & 0.00051 & 0.00051 \\
p-value (Slope) & 0.143 & 0.059 & 0.006 & 0.003 \\
Intercept & 0.4082 & 0.4334 & 0.3967 & 0.3889 \\
SE (Intercept) & 0.0279 & 0.0261 & 0.0247 & 0.0252 \\
\textbf{\(WLS\)} & & & & \\
Slope & 0.00183 & 0.00179 & 0.00205 & 0.00216 \\
SE (Slope) & 0.00054 & 0.00045 & 0.00051 & 0.00051 \\
p-value (Slope) & \textless{} 0.001 & \textless{} 0.001 & \textless{}
0.001 & \textless{} 0.001 \\
Intercept & 0.3157 & 0.3746 & 0.3356 & 0.3238 \\
SE (Intercept) & 0.0329 & 0.0215 & 0.0305 & 0.0348 \\
\textbf{\(GLS\)} & & & & \\
Slope & 0.00196 & 0.00176 & 0.00213 & 0.00225 \\
SE (Slope) & 0.00057 & 0.00046 & 0.00055 & 0.00055 \\
p-value (Slope) & \textless{} 0.001 & \textless{} 0.001 & \textless{}
0.001 & \textless{} 0.001 \\
Intercept & 0.3033 & 0.3766 & 0.3297 & 0.3165 \\
SE (Intercept) & 0.0337 & 0.0219 & 0.0310 & 0.0357 \\
\textbf{\(MEM\)} & & & & \\
Slope & 0.00081 & 0.00097 & 0.00134 & 0.00145 \\
SE (Slope) & 0.00011 & 0.00008 & 0.00010 & 0.00010 \\
p-value (Slope) & \textless{} 0.001 & \textless{} 0.001 & \textless{}
0.001 & \textless{} 0.001 \\
Intercept & 0.4134 & 0.4401 & 0.4015 & 0.3935 \\
SE (Intercept) & 0.0057 & 0.0057 & 0.0056 & 0.0055 \\
\end{longtable}

\section{Robustness to the LULC denominator: the GCB bookkeeping
mean}\label{robustness-to-the-lulc-denominator-the-gcb-bookkeeping-mean}

The headline analysis builds the single airborne-fraction series on the
mean of all \(69\) \(LULC\) definitions. A natural question is whether
the positive trend depends on that expanded ensemble or survives when
the denominator is the conventional Global Carbon Budget land-use
series, the mean of the three bookkeeping models (\(BLUE\), \(OSCAR\),
\(LUCE\)). Table~\ref{tbl-gcb-mean} re-estimates the
\(OLS\)/\(WLS\)/\(GLS\) ladder with this GCB bookkeeping mean as the
\(AF\) denominator, under the same preferred \(RONI\) (lagged one year)
+ \(SAOD\) specification. Only the point-estimate denominator is
changed: the delta-method measurement-error weights still use the full
\(69\)-panel cross-series dispersion
\(\widehat{\operatorname{Var}}(\hat{C}_t)\), so the table isolates the
effect of the conventional denominator choice while retaining the
ensemble-derived weighting. The \(MEM\) is omitted because it is
inherently the \(69\)-series ensemble estimator.

The conclusion is unchanged. Under \(WLS\) and \(GLS\) the trend is
positive and significant in every growth specification (slopes
\(\approx 0.0024\)--\(0.0028\) per year, all \(p<0.001\)), including the
unadjusted series, and the smoothed-\(SAOD\) point estimates
(\(\approx 0.0026\)--\(0.0028\)) are close to the \(69\) \(LULC\)-mean
headline values. \(OLS\) on the GCB mean is again the weakest estimator,
with larger \(p\)-values and more sensitivity to endpoint exclusion;
re-estimating on the sample ending in \(2023\) leaves every \(WLS\) and
\(GLS\) column significant at \(p<0.001\). The positive \(AF\) trend
detected by the weighted single-series estimators is therefore robust to
using the conventional GCB \(LULC\) mean in place of the \(69\) \(LULC\)
mean.

\begin{longtable}[]{@{}
  >{\raggedright\arraybackslash}p{(\linewidth - 8\tabcolsep) * \real{0.2593}}
  >{\raggedleft\arraybackslash}p{(\linewidth - 8\tabcolsep) * \real{0.1852}}
  >{\raggedleft\arraybackslash}p{(\linewidth - 8\tabcolsep) * \real{0.1852}}
  >{\raggedleft\arraybackslash}p{(\linewidth - 8\tabcolsep) * \real{0.1852}}
  >{\raggedleft\arraybackslash}p{(\linewidth - 8\tabcolsep) * \real{0.1852}}@{}}
\caption{\(AF\) trend slope by estimator and growth specification, full
sample (\(1959\)--\(2024\)), with the airborne-fraction series built on
the GCB bookkeeping-model mean as the denominator. The delta-method
weights use the full \(69\)-panel cross-series dispersion. Columns and
standard-error conventions are as in the estimator ladder table of the
main text. Full results, including intercepts, standard errors, and the
\(1959\)--\(2023\) subsample, are available in the GitHub
repository.}\label{tbl-gcb-mean}\tabularnewline
\toprule\noalign{}
\begin{minipage}[b]{\linewidth}\raggedright
\(AF\) trend slope (per year)
\end{minipage} & \begin{minipage}[b]{\linewidth}\raggedleft
\(AF^{RAW}\)
\end{minipage} & \begin{minipage}[b]{\linewidth}\raggedleft
\(AF^{ADJ}\) (SAOD)
\end{minipage} & \begin{minipage}[b]{\linewidth}\raggedleft
\(AF^{ADJ}\) (SAOD*, full)
\end{minipage} & \begin{minipage}[b]{\linewidth}\raggedleft
\(AF^{ADJ}\) (SAOD*, excl.)
\end{minipage} \\
\midrule\noalign{}
\endfirsthead
\toprule\noalign{}
\begin{minipage}[b]{\linewidth}\raggedright
\(AF\) trend slope (per year)
\end{minipage} & \begin{minipage}[b]{\linewidth}\raggedleft
\(AF^{RAW}\)
\end{minipage} & \begin{minipage}[b]{\linewidth}\raggedleft
\(AF^{ADJ}\) (SAOD)
\end{minipage} & \begin{minipage}[b]{\linewidth}\raggedleft
\(AF^{ADJ}\) (SAOD*, full)
\end{minipage} & \begin{minipage}[b]{\linewidth}\raggedleft
\(AF^{ADJ}\) (SAOD*, excl.)
\end{minipage} \\
\midrule\noalign{}
\endhead
\bottomrule\noalign{}
\endlastfoot
\textbf{OLS}: GCB mean, unweighted & & & & \\
Slope & 0.00158 & 0.00177 & 0.00205 & 0.00214 \\
SE (Slope) & 0.00062 & 0.00062 & 0.00051 & 0.00052 \\
p-value (Slope) & 0.011 & 0.004 & \textless{} 0.001 & \textless{}
0.001 \\
\textbf{WLS}: GCB mean, dispersion weighted & & & & \\
Slope & 0.00238 & 0.00242 & 0.00258 & 0.00266 \\
SE (Slope) & 0.00053 & 0.00053 & 0.00052 & 0.00051 \\
p-value (Slope) & \textless{} 0.001 & \textless{} 0.001 & \textless{}
0.001 & \textless{} 0.001 \\
\textbf{GLS}: as \(WLS\), with \(AR(1)\) errors & & & & \\
Slope & 0.00254 & 0.00242 & 0.00266 & 0.00276 \\
SE (Slope) & 0.00057 & 0.00054 & 0.00057 & 0.00055 \\
p-value (Slope) & \textless{} 0.001 & \textless{} 0.001 & \textless{}
0.001 & \textless{} 0.001 \\
\end{longtable}

\section{Robustness to the peat
augmentation}\label{robustness-to-the-peat-augmentation}

The bookkeeping series (\(BLUE\), \(OSCAR\), \(LUCE\)) include peatland
emissions by construction, but the \(22\) process-based land models do
not. To make them comparable we augment each process model by a peat
component, and because three peat products are available (FAO\_peat,
LPX\_Bern\_peat, ORCHIDEE\_peat) we form all three variants of each
process model (Methods). A by-product of this construction is that each
of the \(22\) process models enters the \(69\)-series panel three times,
once per peat product, raising the reasonable concern that the headline
trend could be an artifact of using some of the underlying data three
times. Two considerations show that it is not.

First, the peat product is not what distinguishes the series: the three
peat estimates are small (\(\approx 0.2\) GtC yr\(^{-1}\), against a
total \(LULC\) flux of \(\approx 1.5\) GtC yr\(^{-1}\)) and mutually
similar, so which one is added barely moves a process model.
Table~\ref{tbl-peat} lower block quantifies this: the process-model
ensemble mean sits about \(0.3\) \(GtC\) below the bookkeeping reference
regardless of the peat product, and the spread between the three
products (\(\approx 0.05\) GtC in bias) is an order of magnitude smaller
than that process-versus-bookkeeping gap. There is accordingly no
statistically preferred peat variant; ORCHIDEE\_peat is marginally
closest to the bookkeeping mean in level (it is the largest of the
three, so it closes most of the gap) and LPX\_Bern\_peat tracks its
year-to-year shape best, but the differences are immaterial.

Second, collapsing the triple counting leaves the result unchanged.
Table~\ref{tbl-peat} re-estimates the full estimator ladder on three
reduced \(25\)-series panels, each retaining the three bookkeeping
series and augmenting the \(22\) process models by a single peat
product, alongside the full \(69\)-series headline. Under the preferred
\(RONI\) (lagged one year) + \(SAOD\) specification the \(AF\) trend is
positive and significant across \(OLS\), \(WLS\), \(GLS\), and \(MEM\),
for every peat choice and in both the full sample and the sample ending
in \(2023\). The full-\(69\) estimate is essentially indistinguishable
from the three single-peat panels (\(OLS\)
\(\approx 0.0016\)--\(0.0017\), \(GLS\) \(\approx 0.0024\) per year),
and if anything the single-peat panels give marginally larger slopes.
The positive trend is therefore a property of the ensemble mean and its
dispersion-based weighting, not of the threefold representation of the
process models.

\begin{longtable}[]{@{}
  >{\raggedright\arraybackslash}p{(\linewidth - 8\tabcolsep) * \real{0.2593}}
  >{\raggedleft\arraybackslash}p{(\linewidth - 8\tabcolsep) * \real{0.1852}}
  >{\raggedleft\arraybackslash}p{(\linewidth - 8\tabcolsep) * \real{0.1852}}
  >{\raggedleft\arraybackslash}p{(\linewidth - 8\tabcolsep) * \real{0.1852}}
  >{\raggedleft\arraybackslash}p{(\linewidth - 8\tabcolsep) * \real{0.1852}}@{}}
\caption{\(AF\) trend slope by \(LULC\) panel and estimator (upper
block), preferred \(RONI\) (lag-1) + \(SAOD\) specification with
Pinatubo \(1991\)--\(93\) smoothed to the full-series mean. Each
\(25\)-series panel keeps the three bookkeeping series and augments the
\(22\) process models by a single peat product. \(OLS\)/\(WLS\)/\(GLS\)
use \(HAC\) standard errors; \(MEM\) is the \(AR(1)\) mixed-effects
model. The lower block reports the bias, RMSE, and correlation of each
single-peat process-model mean against the bookkeeping
(\(BLUE\)/\(OSCAR\)/\(LUCE\)) mean.}\label{tbl-peat}\tabularnewline
\toprule\noalign{}
\begin{minipage}[b]{\linewidth}\raggedright
Panel / Estimator
\end{minipage} & \begin{minipage}[b]{\linewidth}\raggedleft
Slope (1959--2024)
\end{minipage} & \begin{minipage}[b]{\linewidth}\raggedleft
p (1959--2024)
\end{minipage} & \begin{minipage}[b]{\linewidth}\raggedleft
Slope (1959--2023)
\end{minipage} & \begin{minipage}[b]{\linewidth}\raggedleft
p (1959--2023)
\end{minipage} \\
\midrule\noalign{}
\endfirsthead
\toprule\noalign{}
\begin{minipage}[b]{\linewidth}\raggedright
Panel / Estimator
\end{minipage} & \begin{minipage}[b]{\linewidth}\raggedleft
Slope (1959--2024)
\end{minipage} & \begin{minipage}[b]{\linewidth}\raggedleft
p (1959--2024)
\end{minipage} & \begin{minipage}[b]{\linewidth}\raggedleft
Slope (1959--2023)
\end{minipage} & \begin{minipage}[b]{\linewidth}\raggedleft
p (1959--2023)
\end{minipage} \\
\midrule\noalign{}
\endhead
\bottomrule\noalign{}
\endlastfoot
\textbf{Full (69), headline} & & & & \\
\(OLS\) & 0.00164 & 0.002 & 0.0014 & 0.006 \\
\(WLS\) & 0.00227 & \textless{} 0.001 & 0.00205 & \textless{} 0.001 \\
\(GLS\) & 0.00237 & \textless{} 0.001 & 0.00213 & \textless{} 0.001 \\
\(MEM\) & 0.00159 & \textless{} 0.001 & 0.00134 & \textless{} 0.001 \\
\textbf{FAO\_peat (25)} & & & & \\
\(OLS\) & 0.00161 & 0.002 & 0.00137 & 0.006 \\
\(WLS\) & 0.00226 & \textless{} 0.001 & 0.00204 & \textless{} 0.001 \\
\(GLS\) & 0.00236 & \textless{} 0.001 & 0.00212 & \textless{} 0.001 \\
\(MEM\) & 0.00156 & \textless{} 0.001 & 0.00131 & \textless{} 0.001 \\
\textbf{LPX\_Bern\_peat (25)} & & & & \\
\(OLS\) & 0.00172 & 0.001 & 0.00148 & 0.004 \\
\(WLS\) & 0.0023 & \textless{} 0.001 & 0.00209 & \textless{} 0.001 \\
\(GLS\) & 0.0024 & \textless{} 0.001 & 0.00216 & \textless{} 0.001 \\
\(MEM\) & 0.00167 & \textless{} 0.001 & 0.00142 & \textless{} 0.001 \\
\textbf{ORCHIDEE\_peat (25)} & & & & \\
\(OLS\) & 0.0017 & 0.001 & 0.00146 & 0.004 \\
\(WLS\) & 0.00231 & \textless{} 0.001 & 0.0021 & \textless{} 0.001 \\
\(GLS\) & 0.00241 & \textless{} 0.001 & 0.00217 & \textless{} 0.001 \\
\(MEM\) & 0.00165 & \textless{} 0.001 & 0.0014 & \textless{} 0.001 \\
& & & & \\
\textbf{Peat product} & \textbf{Bias (GtC)} & \textbf{RMSE (GtC)} &
\textbf{Correlation} & \\
FAO\_peat & -0.3129 & 0.3493 & 0.7017 & \\
LPX\_Bern\_peat & -0.3251 & 0.3518 & 0.8089 & \\
ORCHIDEE\_peat & -0.2708 & 0.304 & 0.7701 & \\
\end{longtable}

\section{Robustness to fossil-emission
uncertainty}\label{robustness-to-fossil-emission-uncertainty}

The total-emissions variance that drives the delta-method weights
(Methods) adds a fossil-emission term \((\kappa\,FF_t)^2\) to the
\(LULC\) dispersion, with \(\kappa\) the fractional \(1\sigma\)
uncertainty on fossil CO\(_2\). The headline analysis uses the Global
Carbon Budget's reported value of \(\kappa=0.05\)
\autocite{Friedlingstein2025}. Table~\ref{tbl-ff-sensitivity}
re-estimates the second-stage \(WLS\) and \(GLS\) \(AF\)-trend slopes
for \(\kappa\in\{0,0.03,0.05,0.07\}\), where \(\kappa=0\) recovers the
\(LULC\)-only dispersion weighting. Increasing \(\kappa\) attenuates the
weighted slopes modestly and near-monotonically (most visibly for the
unadjusted series, whose weights are purely denominator-driven) but
every \(WLS\) and \(GLS\) estimate remains positive and significant at
\(p<0.001\) across the full range. Fossil-emission uncertainty therefore
tightens the estimates slightly without changing any qualitative
conclusion. \(OLS\) and \(MEM\) do not use these weights.

\begin{longtable}[]{@{}
  >{\raggedright\arraybackslash}p{(\linewidth - 8\tabcolsep) * \real{0.2000}}
  >{\raggedleft\arraybackslash}p{(\linewidth - 8\tabcolsep) * \real{0.2000}}
  >{\raggedleft\arraybackslash}p{(\linewidth - 8\tabcolsep) * \real{0.2000}}
  >{\raggedleft\arraybackslash}p{(\linewidth - 8\tabcolsep) * \real{0.2000}}
  >{\raggedleft\arraybackslash}p{(\linewidth - 8\tabcolsep) * \real{0.2000}}@{}}
\caption{AF trend slope (per year), full sample (\(1959\)--\(2024\)),
preferred RONI (lag-1) + SAOD specification, as the assumed
fossil-emission \(1\sigma\) uncertainty \(\kappa\) varies. The
delta-method weights use
\(\operatorname{Var}(\hat C_t) = \operatorname{Var}_j(FF_t + LULC_t^{(j)}) + (\kappa\,FF_t)^2\);
\(\kappa=0\) is the LULC-only weighting and \(\kappa=0.05\) is the
headline. Every entry is significant at \(p<0.001\) (HAC standard
errors); full detail in
\protect\texttt{results/ff\_uncertainty\_sensitivity.md}.}\label{tbl-ff-sensitivity}\tabularnewline
\toprule\noalign{}
\begin{minipage}[b]{\linewidth}\raggedright
\(\kappa\) (fossil \(1\sigma\))
\end{minipage} & \begin{minipage}[b]{\linewidth}\raggedleft
\(AF^{RAW}\)
\end{minipage} & \begin{minipage}[b]{\linewidth}\raggedleft
\(AF^{ADJ}\) (\(SAOD\))
\end{minipage} & \begin{minipage}[b]{\linewidth}\raggedleft
\(AF^{ADJ}\) (\(SAOD^*\), full)
\end{minipage} & \begin{minipage}[b]{\linewidth}\raggedleft
\(AF^{ADJ}\) (\(SAOD^*\), excl.)
\end{minipage} \\
\midrule\noalign{}
\endfirsthead
\toprule\noalign{}
\begin{minipage}[b]{\linewidth}\raggedright
\(\kappa\) (fossil \(1\sigma\))
\end{minipage} & \begin{minipage}[b]{\linewidth}\raggedleft
\(AF^{RAW}\)
\end{minipage} & \begin{minipage}[b]{\linewidth}\raggedleft
\(AF^{ADJ}\) (\(SAOD\))
\end{minipage} & \begin{minipage}[b]{\linewidth}\raggedleft
\(AF^{ADJ}\) (\(SAOD^*\), full)
\end{minipage} & \begin{minipage}[b]{\linewidth}\raggedleft
\(AF^{ADJ}\) (\(SAOD^*\), excl.)
\end{minipage} \\
\midrule\noalign{}
\endhead
\bottomrule\noalign{}
\endlastfoot
\textbf{WLS} & & & & \\
\(0\) (\(LULC\) only) & 0.00230 & 0.00204 & 0.00239 & 0.00253 \\
\(3\%\) & 0.00218 & 0.00201 & 0.00234 & 0.00246 \\
\(5\%\) (headline) & 0.00204 & 0.00199 & 0.00227 & 0.00238 \\
\(7\%\) & 0.00192 & 0.00199 & 0.00220 & 0.00229 \\
\textbf{GLS} & & & & \\
\(0\) (LULC only) & 0.00252 & 0.00203 & 0.00250 & 0.00266 \\
\(3\%\) & 0.00237 & 0.00199 & 0.00244 & 0.00259 \\
\(5\%\) (headline) & 0.00221 & 0.00197 & 0.00237 & 0.00249 \\
\(7\%\) & 0.00206 & 0.00197 & 0.00230 & 0.00240 \\
\end{longtable}

\section{ENSO lag selection}\label{enso-lag-selection}

\begin{figure}

\centering{

\pandocbounded{\includegraphics[keepaspectratio]{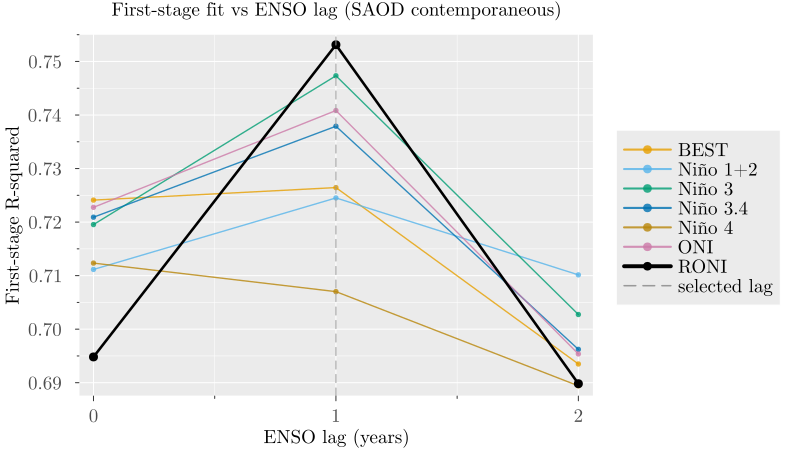}}

}

\caption{\label{fig-lag-selection}First-stage \(R^2\) against ENSO lag
for each of the seven ENSO indices (volcanic index SAOD held
contemporaneous). RONI (black) shows the largest gain from lag zero to
lag one, the selected lag.}

\end{figure}

In the natural-variability specification (the first-stage growth
regression in Methods), the \(ENSO\) index enters with a one-year lag
and the volcanic index contemporaneously. We select the \(ENSO\) lag on
the preferred index, \(RONI\), by first-stage fit, sweeping lags
\(0\)--\(2\) years for the \(ENSO\) regressor and \(0\)--\(1\) years for
the volcanic regressor on the full \(1959\)--\(2024\) sample, and report
whether the same lag also improves fit for the other ENSO indices
considered in the sensitivity grid.

\(RONI\)'s first-stage \(R^2\) rises from \(0.6948\) at lag zero to
\(0.7531\) at lag one before falling to \(0.6898\) at lag two. The
increase is the largest gain of any index, so we adopt lag one. This
choice is either optimal or fit-improving for six of the seven \(ENSO\)
indices considered (Figure~\ref{fig-lag-selection},
Table~\ref{tbl-lag-selection}); only Niño 4 marginally prefers the
contemporaneous specification. Holding \(ENSO\) at \(RONI\) lagged one
year, the volcanic index is best contemporaneous: \(SAOD\)'s first-stage
\(R^2\) is \(0.7531\) at lag zero versus \(0.7114\) at lag one. The
contemporaneous fit is clearly superior, consistent with the documented
near-immediate effect of stratospheric aerosol loading on the
terrestrial and ocean carbon flux \autocite{Betts2016}.

\begin{longtable}[]{@{}
  >{\raggedright\arraybackslash}p{(\linewidth - 8\tabcolsep) * \real{0.3333}}
  >{\raggedleft\arraybackslash}p{(\linewidth - 8\tabcolsep) * \real{0.1667}}
  >{\raggedleft\arraybackslash}p{(\linewidth - 8\tabcolsep) * \real{0.1667}}
  >{\raggedleft\arraybackslash}p{(\linewidth - 8\tabcolsep) * \real{0.1667}}
  >{\raggedleft\arraybackslash}p{(\linewidth - 8\tabcolsep) * \real{0.1667}}@{}}
\caption{First-stage \(R^2\) at each \(ENSO\) lag, volcanic index
(\(SAOD\)) held
contemporaneous.}\label{tbl-lag-selection}\tabularnewline
\toprule\noalign{}
\begin{minipage}[b]{\linewidth}\raggedright
\(ENSO\) index
\end{minipage} & \begin{minipage}[b]{\linewidth}\raggedleft
\(R^2\) (lag 0)
\end{minipage} & \begin{minipage}[b]{\linewidth}\raggedleft
\(R^2\) (lag 1)
\end{minipage} & \begin{minipage}[b]{\linewidth}\raggedleft
\(R^2\) (lag 2)
\end{minipage} & \begin{minipage}[b]{\linewidth}\raggedleft
Best lag
\end{minipage} \\
\midrule\noalign{}
\endfirsthead
\toprule\noalign{}
\begin{minipage}[b]{\linewidth}\raggedright
\(ENSO\) index
\end{minipage} & \begin{minipage}[b]{\linewidth}\raggedleft
\(R^2\) (lag 0)
\end{minipage} & \begin{minipage}[b]{\linewidth}\raggedleft
\(R^2\) (lag 1)
\end{minipage} & \begin{minipage}[b]{\linewidth}\raggedleft
\(R^2\) (lag 2)
\end{minipage} & \begin{minipage}[b]{\linewidth}\raggedleft
Best lag
\end{minipage} \\
\midrule\noalign{}
\endhead
\bottomrule\noalign{}
\endlastfoot
\(BEST\) & 0.7241 & 0.7264 & 0.6935 & 1 \\
Niño 1+2 & 0.7112 & 0.7245 & 0.7101 & 1 \\
Niño 3 & 0.7195 & 0.7473 & 0.7027 & 1 \\
Niño 3.4 & 0.7209 & 0.7379 & 0.6962 & 1 \\
Niño 4 & 0.7123 & 0.7070 & 0.6894 & 0 \\
\(ONI\) & 0.7228 & 0.7408 & 0.6954 & 1 \\
\(RONI\) & 0.6948 & 0.7531 & 0.6898 & 1 \\
\end{longtable}

\section{Theoretical Background}\label{theoretical-background}

This section collects the formal derivations behind the estimators
summarised in Methods: the first-stage natural-variability weighting,
the delta-method variance of the airborne-fraction ratio, the two-stage
\(WLS\)/\(GLS\) trend regression, the \(MEM\) cross-check, and the
carbon-budget scaling. The Monte Carlo study that motivates the choice
of estimator closes the section.

\subsection{First-stage weighting and the fitted-growth
variance}\label{first-stage-weighting-and-the-fitted-growth-variance}

The natural-variability regression of the main text is given by:
\begin{equation}\protect\phantomsection\label{si-eq-growth-reg}{
\hat{G}_t = \gamma_0 + \gamma_1 \hat{C}_t + \gamma_2 \,\text{ENSO}_{t-1} + \gamma_3 \,\text{SAOD}_t + \varepsilon_t,
}\end{equation}

where the total-emissions regressor is the sum of fossil emissions and
the LULC-panel mean, \[
\hat C_t = FF_t + \overline{LULC}_t, \qquad \overline{LULC}_t = \frac{1}{n_t}\sum_{j=1}^{n_t} LULC_{tj},
\] with \(n_t=3\) for the GCB bookkeeping mean and \(n_t=69\) for the
all-LULC mean. To account for measurement uncertainty in total emissions
we combine two independent components: the cross-measurement dispersion
of the LULC panel \autocite{Cochran1977} and the Global Carbon Budget's
reported \(\pm5\%\) (\(1\sigma\)) uncertainty on fossil emissions
\autocite{Friedlingstein2025}.

Propagating \(\hat{C}_t\) through the linear model by the delta method
gives the dispersion contribution
\(\gamma_1^2\,\widehat{\operatorname{Var}}(\hat{C}_t)\) to the variance
of growth. The growth observation also carries idiosyncratic noise with
variance \(\sigma^2\) (the residual variance), so the total conditional
variance used for weighting is the sum of the two independent
components, \[
\omega_t^2 = \sigma^2 + \gamma_1^2\,\widehat{\operatorname{Var}}(\hat{C}_t),
\] and the first-stage WLS weights are its reciprocal. The \(GLS\)
variant adds an \(AR(1)\) error structure with the same controls and
weights. The natural-variability-adjusted growth series subtracts only
the estimated ENSO and volcanic contributions, \[
\hat{G}_t^{ADJ} = G_t - \hat{\gamma}_2 \,\text{ENSO}_{t-1} - \hat{\gamma}_3 \,\text{SAOD}_t,
\] with coefficients taken from the fit, so the emissions-driven
component and the residual are retained.

\subsection{Delta method for the ratio
variance}\label{delta-method-for-the-ratio-variance}

Our second-stage estimator propagates denominator measurement
uncertainty into the airborne-fraction trend through a delta-method
variance for the ratio \(AF_t = \hat{G}_t^{ADJ}/\hat{C}_t\)
\autocite{Fuller1987,Carroll2006,Oehlert1992,Aitken1935}. Define the
random vector and the function of interest as \[
X_t=
\begin{bmatrix}
\hat{G}_t\\
\hat C_t
\end{bmatrix},
\qquad
g(X_t)=\frac{\hat{G}_t}{\hat C_t},
\] and let \[
\sigma_{G,t}^2 = \operatorname{Var}(\hat{G}_t),\qquad \sigma_{C,t}^2 = \operatorname{Var}(\hat C_t),\qquad \sigma_{GC,t} = \operatorname{Cov}(\hat{G}_t,\hat C_t).
\]

Linearising \(g\) around the mean vector
\((\mu_{G,t},\mu_{C,t})=\mathbb E[(\hat{G}_t,\hat C_t)]\) and applying
the delta-method variance formula
\(\operatorname{Var}(g(X_t))\approx \nabla g^\top\Sigma_t\nabla g\) with
gradient \(\nabla g(G,C)=(1/C,\,-G/C^2)^\top\) and covariance
\(\Sigma_t=\bigl[\begin{smallmatrix}\sigma_{G,t}^2 & \sigma_{GC,t}\\ \sigma_{GC,t} & \sigma_{C,t}^2\end{smallmatrix}\bigr]\)
gives, with plug-in evaluation at \((G_t,C_t)\), \[
\operatorname{Var}(AF_t)\approx\left(\frac{1}{C_t}\right)^2 \sigma_{G,t}^2+\left(\frac{G_t}{C_t^2}\right)^2 \sigma_{C,t}^2-2\frac{G_t}{C_t^3}\sigma_{GC,t}.
\]

In practice we use plug-in estimates
\(\sigma^2_{G,t}\approx\widehat{\operatorname{Var}}(\hat{G}_t)\) and
\(\sigma^2_{C,t}\approx\widehat{\operatorname{Var}}(\hat C_t)\), and set
the covariance term \(\sigma_{GC,t}\approx 0\). The two uncertainties
have distinct sources, which is what makes this negligible rather than
merely convenient. The numerator's variance derives entirely from the
estimated first-stage coefficients \(\hat\gamma_2,\hat\gamma_3\), since
the observed growth \(G_t\) enters \(\hat{G}_t^{ADJ}\) as data and
carries no estimation error; the denominator's derives from cross-model
measurement error in \(\hat C_t\). The two are linked only because
\(\hat C_t\) supplies one of the \(n=66\) rows of the first-stage design
from which \(\hat\gamma\) is estimated, so the induced covariance enters
at \(O(n^{-1})\) relative to either variance term. Dropping it yields
the variance proxy used in the main text, \[
\widehat{\operatorname{Var}}(AF_t) = \left(\frac{1}{C_t}\right)^2 \widehat{\operatorname{Var}}(\hat{G}_t) + \left(\frac{G_t}{C_t^2}\right)^2 \widehat{\operatorname{Var}}(\hat C_t).
\]

Here the numerator variance is the sampling variance of the adjusted
growth series
\(\hat{G}_t^{ADJ} = G_t - \hat\gamma_2\,\text{ENSO}_{t-1} - \hat\gamma_3\,\text{SAOD}_t\).
Only the two subtracted coefficients carry estimation uncertainty into
\(\hat{G}_t^{ADJ}\): the observed growth \(G_t\) is data, and
\(\hat\gamma_0\) and \(\hat\gamma_1\) never enter the adjusted series.
Writing \(\widehat{\operatorname{Cov}}(\hat\gamma)\) for the first-stage
coefficient covariance, and letting
\(z_t=(0,0,\text{ENSO}_{t-1},\text{SAOD}_t)^\top\) be the first-stage
design row of Equation~\ref{si-eq-growth-reg} restricted to the
natural-variability columns, this is \[
\widehat{\operatorname{Var}}(\hat{G}_t^{ADJ}) = z_t^\top\,\widehat{\operatorname{Cov}}(\hat\gamma)\,z_t .
\] Because the natural-variability component subtracted from growth uses
the \(GLS\) (Prais--Winsten, first-stage-weighted) coefficients,
\(\widehat{\operatorname{Cov}}(\hat\gamma)\) is the coefficient
covariance of that same \(GLS\) fit; using the unweighted \(OLS\)
covariance \(\hat\sigma^2(X^\top X)^{-1}\) instead would propagate the
uncertainty of coefficients that were never used to form
\(\hat{G}_t^{ADJ}\). Restricting the quadratic form to \(z_t\) rather
than the full design row
\(x_t=(1,\hat C_t,\text{ENSO}_{t-1},\text{SAOD}_t)^\top\) matters
substantively as well as formally:
\(x_t^\top\widehat{\operatorname{Cov}}(\hat\gamma)\,x_t\) is the
variance of the fitted value \(x_t^\top\hat\gamma\), a different object,
and because the \(\hat C_t\) column is an order of magnitude larger than
the others and trends upward over the sample, it would dominate the
quadratic form and impose a spurious upward trend on the second-stage
weights. This is the object that propagates the first-stage estimation
uncertainty into the airborne-fraction trend and its standard errors,
and it is distinct from the first-stage weighting variance
\(\omega_t^2\) above. For the unadjusted \(AF^{RAW}\) series there is no
first stage, so \(\widehat{\operatorname{Var}}(\hat{G}_t)=0\) and only
the denominator term contributes.

\subsection{Two-stage WLS/GLS trend
regression}\label{two-stage-wlsgls-trend-regression}

With the point estimate \(AF_t\) and its variance proxy in hand, we fit
the linear trend \[
AF_t = \alpha + \beta t + \varepsilon_t, \quad \varepsilon_t \sim (0, \sigma_t^2), \quad \sigma_t^2 = \operatorname{Var}(AF_t),
\] by \(WLS\) with weights \(w_t=1/\sigma_t^2\), so that years with less
dispersed \(LULC\) measurements, or better-measured denominator, receive
greater weight. Testing for a trend is a test of \(\beta=0\) against
\(\beta\neq 0\), and \(WLS\) uses the multiple information entirely
through this delta-method weighting scheme. A \(GLS\) variant adds a
Prais--Winsten \(AR(1)\) error structure \autocite{PraisWinsten1954}.
The regression estimators use heteroskedasticity- and
autocorrelation-consistent (\(HAC\)) standard errors
\autocite{NeweyWest1987,Andrews1991}. For the unadjusted series
(\(AF^{RAW}\)) the growth series is observed rather than estimated, so
\(\operatorname{Var}(G_t)=0\) in the delta-method variance and the
weights propagate denominator measurement uncertainty alone. The
\(WLS\)/\(GLS\) slopes differ from the \(MEM\) slope because they
operate on a single weighted \(AF\) series, whereas the \(MEM\) pools
the full \(69\)-series panel.

\subsection{Mixed-effects model}\label{mixed-effects-model}

As a cross-check on the single-series estimators, we fit a linear
mixed-effects model (\(MEM\)) to the panel of yearly \(AF\) values
constructed from each \(LULC\) measurement series. For each year \(t\)
and series \(j\), we define the airborne fraction as: \[
AF_{t,j} = \frac{\hat{G}_t^{ADJ}}{FF_t + LULC_{t,j}},
\] and we estimate: \[
AF_{t,j} = (\mu_\alpha + u_{\alpha j}) + (\mu_\beta + u_{\beta j})\,t + \varepsilon_{t,j},
\] with random effects and \(AR(1)\) errors \[
\begin{bmatrix}u_{\alpha j} \\ u_{\beta j}\end{bmatrix}
\sim N\!\left(
\begin{bmatrix}0\\0\end{bmatrix},
\Sigma_u
\right),
\quad
\varepsilon_{t,j} = \rho \varepsilon_{t-1,j} + \eta_{t,j}, \quad \eta_{t,j} \sim N(0,\sigma^2).
\]

Each \(LULC\) series thus has its own level and trend while the model
estimates a common population-average trend \(\mu_\beta\), the estimand
of interest, interpreted as the average annual change in AF across the
LULC definitions. The model is estimated by restricted maximum
likelihood \autocite{Henderson1953,Bolker2009} with model-based standard
errors and Wald tests. Identification relies on standard \(MEM\)
conditions: conditional linearity in time, random effects centred at
zero and independent across definitions, and mean-zero residuals
conditional on the fixed and random effects. Estimating the model
without the \(AR(1)\) structure gives qualitatively similar results.

\subsection{Carbon-budget scaling}\label{carbon-budget-scaling}

To illustrate the implications of a rising \(AF\) for the remaining
carbon budget, we use a simple scaling argument. Let \(B\) denote the
remaining carbon budget for a given temperature target under a
counterfactual constant airborne fraction \(\overline{AF}\). If the
relevant path-weighted airborne fraction over the remaining emissions
trajectory is \(AF^{*} > \overline{AF}\), the budget consistent with the
same atmospheric \(CO_2\) accumulation, and thus the same temperature
target, is approximately \(B^{*} \approx B\,\overline{AF}/AF^{*}\),
implying an effective budget reduction \[
\Delta B \approx B \left(1 - \frac{\overline{AF}}{AF^{*}}\right).
\]

We take \(B = 300\)--\(400\) \(GtCO_2\), the IPCC AR6 remaining carbon
budget for \(1.5\) \(^\circ\)C at \(50\%\) probability assessed from the
start of \(2023\) \autocite{IPCC_AR6_WG1}, and the reference
\(\overline{AF} = 0.44\), a commonly reported observational benchmark
under a constant-\(AF\) approximation
\autocite{Canadell2007,Raupach2007,Knorr2009,bennedsenRegressionbasedApproachCO22024}.
For the trend-implied current \(AF\) we use the fitted \(2024\) value
\(\widehat{AF}_{2024} = \widehat{a} + \widehat{b}\times 65\) from the
preferred smoothed specification of the main-text ladder, which spans
\(\widehat{AF}_{2024}\in[0.477,0.500]\): about \(0.48\) for the
measurement-error-weighted single-series estimators (\(WLS\) \(0.477\),
\(GLS\) \(0.477\)) and about \(0.50\) for the unweighted \(OLS\) and
\(MEM\) estimators (\(0.498\) and \(0.500\)). The fractional budget
reduction is then \[
\frac{\Delta B}{B} \approx 1 - \frac{0.440}{\widehat{AF}_{2024}} \in [\,0.077,\ 0.119\,],
\] a reduction of about \(23\)--\(48\) \(GtCO_2\) over the AR6 range, up
to roughly one year of current global fossil fuel \(CO_2\) emissions
\autocite{Friedlingstein2025}.

This calculation rests on two simplifying assumptions. First, it treats
the AR6 budget as if evaluated under a constant airborne fraction
\(\overline{AF}=0.44\); in practice AR6 budgets come from ensembles of
Earth System Models and reduced-complexity models that already project a
gradually rising AF due to carbon-cycle feedbacks, so to the extent
those models embed a rising-\(AF\) trajectory consistent with our
estimates, the incremental budget effect will be smaller
\autocite{IPCC_AR6_WG1}. Second, it uses the endpoint \(AF\) as a single
representative value, whereas the effective reduction depends on the
\(AF\) trajectory over the entire remaining pathway; a full
reconciliation would replace the endpoint with a pathway-weighted \(AF\)
integrated over the remaining emissions trajectory. The calculation is
therefore an illustrative, order-of-magnitude indication of how a rising
\(AF\) tightens the remaining carbon budget, not a precise estimate of
the budget impact.

\subsection{Monte Carlo evidence for multi-source
designs}\label{monte-carlo-evidence-for-multi-source-designs}

A natural concern about the main text's design is whether combining many
noisy, model-dependent \(LULC\) series genuinely improves trend
detection relative to a single conventional \(LULC\) definition.
Moreover, we are interested in which estimator turns the ensemble into
valid inference rather than merely apparent precision. The \(LULC\)
series are highly correlated: they share one atmospheric-growth
numerator and are built from a set of near-identical bookkeeping and
process-model definitions. A Monte Carlo study calibrated to this design
shows that the delta-method \(WLS\)/\(GLS\) estimator converts that
correlated ensemble into a validly quantified, power gain, whereas the
\(MEM\) obtains apparent power by treating correlated series as
independent.

\subsubsection{Design}\label{design}

The main-text estimators regress the airborne fraction on known calendar
time; it is the annual \(AF\) values, not the time axis, that are
uncertain. We therefore generate, for each replication, \(M\)
measurement series of the \(AF\) directly, \[
AF_{i,t} = a + b\,t + \sigma_t\bigl(\sqrt{\rho}\,f_t + \sqrt{1-\rho}\,e_{i,t}\bigr), \qquad f_t, e_{i,t}\sim N(0,1),
\] for \(t=1,\dots,T\) and \(i=1,\dots,M\), with two ingredients
calibrated to the empirical panel. First, the annual reliability
\(\sigma_t\) declines over the record, from \(\sigma_{\text{hi}}=0.11\)
in the earliest years, where the cross-model spread in land-use
emissions is largest, to \(\sigma_{\text{lo}}=0.022\); this is the
reliability profile the delta-method weights encode. Second, the series
share a common annual shock \(f_t\) with correlation \(\rho=0.9\),
representing the shared growth numerator and the
\(\approx\!98\%\)-correlated \(LULC\) definitions, so that averaging or
pooling cannot reduce the noise below the common floor
\(\sigma_t^2\rho\) and additional series do not buy proportional
information. We use \(T=66\) (matching \(1959\)--\(2024\)), \(b=0.001\),
\(a=0.40\), and \(n_{\text{sim}}=10{,}000\) replications for each
\(M\in\{1,3,5,10,20,50,69\}\), the last matching the empirical panel
size.

Because every estimator here regresses an \(AF\)-scale series on time,
the true slope is \(b\) for all of them and estimates are directly
comparable; alongside statistical power (rejection of the no-trend null
at the \(5\%\) level) we therefore report the coverage of the nominal
\(95\%\) confidence interval, a key diagnostic of whether a design's
precision is real. We compare five estimators with the main text's own
inference (\(HAC\) standard errors for the regression estimators):

\begin{enumerate}
\def\labelenumi{\arabic{enumi}.}
\tightlist
\item
  \textbf{Single-series OLS}: \(OLS\) of one randomly drawn series on
  \(t\), mirroring estimation with a single \(LULC\) definition.
\item
  \textbf{Mean OLS}: \(OLS\) of the equal-weight cross-series mean
  \(\overline{AF}_t = M^{-1}\sum_i AF_{i,t}\) on \(t\).
\item
  \textbf{WLS}: \(WLS\) of \(\overline{AF}_t\) on \(t\) with
  delta-method weights
  \(w_t = 1/\widehat{\operatorname{Var}}_i(AF_{i,t})\), the reciprocal
  cross-series variance at time \(t\), mirroring the \(AF\)-trend
  weighting of the main text.
\item
  \textbf{GLS}: the same weighted design with a Prais--Winsten \(AR(1)\)
  error structure.
\item
  \textbf{MEM}: a linear \(MEM\) with random intercepts and slopes by
  series, fitted to the full \(T\times M\) panel, mirroring the panel
  specification of the main text.
\end{enumerate}

\subsubsection{Results}\label{si-results}

\begin{figure}

\centering{

\pandocbounded{\includegraphics[keepaspectratio]{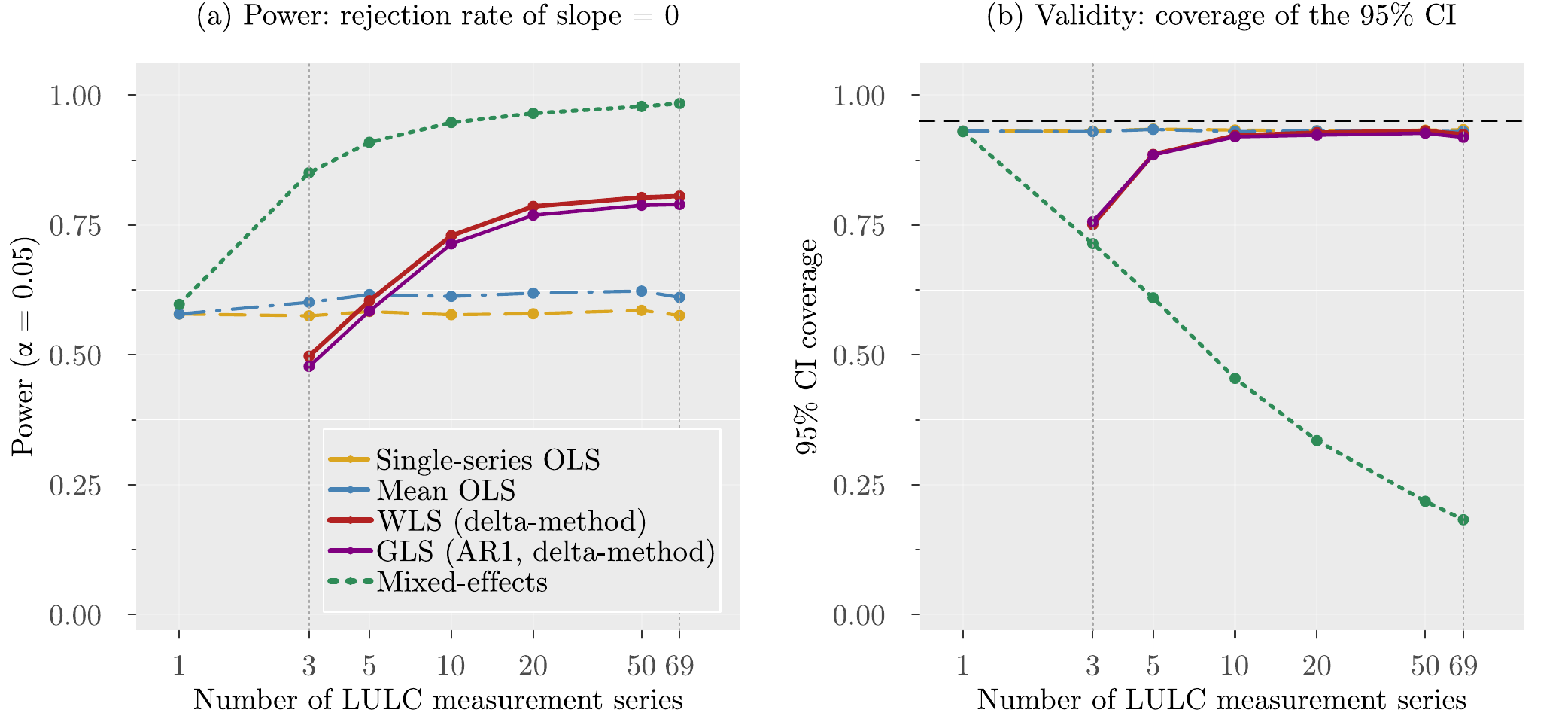}}

}

\caption{\label{fig-mc-power}Monte Carlo comparison of trend estimators
as a function of the number of \(LULC\) measurement series \(M\), over
\(10,000\) replications (\(T=66\), \(b=0.001\), \(a=0.40\), declining
reliability \(\sigma_t: 0.11\to0.022\), cross-series correlation
\(\rho=0.9\)). Panel (a): power, the rejection rate of
\(H_0\colon\text{slope}=0\) at \(\alpha=0.05\). Panel (b): coverage of
the nominal \(95\%\) confidence interval (dashed line). Delta-method
\(WLS\)/\(GLS\) gain power with \(M\) while holding valid coverage; the
\(MEM\) attains higher raw power but its coverage collapses as \(M\)
grows, revealing pseudo-replication. Vertical lines mark \(M=3\) (GCB
bookkeeping mean) and \(M=69\) (full panel); WLS/GLS require
\(M\ge 2\).}

\end{figure}

Table~\ref{tbl-monte-carlo-full} and Figure~\ref{fig-mc-power} report,
for each estimator and \(M\), the Monte Carlo mean slope, mean standard
error, power, and \(95\%\) CI coverage, based on \(10,000\) replications
per cell. \(WLS\) and \(GLS\) are undefined at \(M=1\), where the
cross-series variance cannot be computed.

Every estimator recovers the true slope on average
(\(\overline{\text{slope}}\approx 0.001\) throughout), so the story is
entirely in precision and its validity. Single-series and mean OLS have
flat power in \(M\) (\(\approx 0.58\) and \(\approx 0.61\)): because the
series share the common shock, averaging cannot lower the binding noise.
Both, however, keep valid coverage (\(\approx 0.93\)). The delta-method
\(WLS\) and \(GLS\) estimators exploit the structure of the ensemble,
downweighting the unreliable early decades via the cross-series
dispersion, and their power rises with \(M\) (\(WLS\) from \(0.50\) at
\(M=3\) to \(0.81\) at \(M=69\)) while their coverage returns to the
nominal level once enough series pin down the weights (\(M\gtrsim 10\);
coverage \(\approx 0.92\)--\(0.93\)). At small \(M\) the weights are
poorly estimated and \(WLS\)/\(GLS\) coverage dips (to \(\approx 0.75\)
at \(M=3\)), which is itself an argument for using the full ensemble.
The \(MEM\) attains the highest raw power (\(0.98\) at \(M=69\)), but
its coverage collapses monotonically from \(0.93\) at \(M=1\) to
\(0.18\) at \(M=69\): by treating \(M\) correlated copies of one growth
realization as \(M\) independent series it understates the standard
error by roughly \(\sqrt{M}\) and rejects far too often. Its apparent
power is pseudo-replication, not information.

\begin{longtable}[]{@{}rlrrrr@{}}
\caption{Full Monte Carlo results across \(10,000\) replications, by
estimator and number of measurement series (\(M\)). All estimators
regress an \(AF\)-scale series on time, so the true slope is \(b=0.001\)
throughout and estimates are directly comparable. ``\(95\%\) coverage''
is the fraction of nominal-\(95\%\) confidence intervals containing
\(b\).}\label{tbl-monte-carlo-full}\tabularnewline
\toprule\noalign{}
M & Method & Mean slope & Mean SE & Power & 95\% coverage \\
\midrule\noalign{}
\endfirsthead
\toprule\noalign{}
M & Method & Mean slope & Mean SE & Power & 95\% coverage \\
\midrule\noalign{}
\endhead
\bottomrule\noalign{}
\endlastfoot
1 & Single-series OLS & 0.000998 & 0.000460 & 0.579 & 0.931 \\
1 & Mean OLS & 0.000998 & 0.000460 & 0.579 & 0.931 \\
1 & MEM & 0.000998 & 0.000447 & 0.597 & 0.930 \\
3 & Single-series OLS & 0.000995 & 0.000462 & 0.575 & 0.931 \\
3 & Mean OLS & 0.000996 & 0.000446 & 0.601 & 0.930 \\
3 & WLS & 0.000992 & 0.000586 & 0.498 & 0.751 \\
3 & GLS & 0.000991 & 0.000726 & 0.478 & 0.756 \\
3 & MEM & 0.000996 & 0.000259 & 0.851 & 0.714 \\
5 & Single-series OLS & 0.000997 & 0.000461 & 0.583 & 0.935 \\
5 & Mean OLS & 0.000999 & 0.000443 & 0.616 & 0.934 \\
5 & WLS & 0.000997 & 0.000459 & 0.604 & 0.886 \\
5 & GLS & 0.000996 & 0.000487 & 0.584 & 0.885 \\
5 & MEM & 0.000999 & 0.000201 & 0.909 & 0.610 \\
10 & Single-series OLS & 0.000999 & 0.000462 & 0.577 & 0.933 \\
10 & Mean OLS & 0.001000 & 0.000441 & 0.613 & 0.930 \\
10 & WLS & 0.001000 & 0.000381 & 0.729 & 0.922 \\
10 & GLS & 0.001000 & 0.000391 & 0.714 & 0.920 \\
10 & MEM & 0.001000 & 0.000142 & 0.947 & 0.455 \\
20 & Single-series OLS & 0.001000 & 0.000463 & 0.579 & 0.931 \\
20 & Mean OLS & 0.001001 & 0.000440 & 0.619 & 0.931 \\
20 & WLS & 0.001000 & 0.000355 & 0.786 & 0.928 \\
20 & GLS & 0.001000 & 0.000363 & 0.769 & 0.923 \\
20 & MEM & 0.001001 & 0.000100 & 0.965 & 0.335 \\
50 & Single-series OLS & 0.001001 & 0.000463 & 0.586 & 0.932 \\
50 & Mean OLS & 0.001000 & 0.000439 & 0.623 & 0.930 \\
50 & WLS & 0.001000 & 0.000345 & 0.803 & 0.931 \\
50 & GLS & 0.000999 & 0.000351 & 0.788 & 0.927 \\
50 & MEM & 0.001000 & 0.000064 & 0.978 & 0.218 \\
69 & Single-series OLS & 0.000992 & 0.000461 & 0.576 & 0.933 \\
69 & Mean OLS & 0.000990 & 0.000438 & 0.611 & 0.930 \\
69 & WLS & 0.000994 & 0.000341 & 0.806 & 0.924 \\
69 & GLS & 0.000993 & 0.000347 & 0.790 & 0.919 \\
69 & MEM & 0.000990 & 0.000054 & 0.984 & 0.183 \\
\end{longtable}

This simulation supports the estimation strategy of the main text and
identifies the estimator to trust. Exploiting the full ensemble of
repeated, correlated measurements improves trend detection over a single
definition, but only the delta-method \(WLS\)/\(GLS\) design does so
maintaining valid coverage. The \(MEM\)'s larger raw power is largely an
artifact of pseudo-replication, its confidence intervals undercovering
severely once the panel is large; it is therefore best read as a
corroborating diagnostic rather than the primary inferential estimator.
This mirrors the empirical pattern in the main text, where
single-definition \(OLS\) gives weaker and less robust evidence than the
weighted single-series \(WLS\)/\(GLS\) estimates on the full \(LULC\)
ensemble.

\section{Open Research}\label{si-open-research}

\subsection{Data Availability
Statement}\label{data-availability-statement}

The study uses publicly available Global Carbon Budget \(2025\) data,
accessible at \url{https://globalcarbonbudget.org/gcb-2025/}. The ENSO
indices, including the relative Oceanic Niño Index (RONI) used in the
preferred specification
\autocite{LHeureux2024RONI,Huang_2017_ERSSTv5_JCLI}, were obtained from
the NOAA Physical Sciences Laboratory (\url{https://psl.noaa.gov/enso/},
\url{https://psl.noaa.gov/data/climateindices/}). The stratospheric
aerosol optical depth (SAOD) series used in the preferred specification
is built from the CMIP7 v2.2.1 stratospheric aerosol forcing
\autocite{Aubry2025CMIP7,Aubry2025CMIP7data} (University of Exeter
\texttt{input4MIPs}, \url{https://aims2.llnl.gov/search/input4MIPs/})
for \(1959\)--\(2023\), extended to \(2024\) with GloSSAC v2.24
\autocite{Kovilakam2020GloSSAC,Kovilakam2026GloSSACdata} from the NASA
Atmospheric Science Data Center
(\url{https://asdc.larc.nasa.gov/project/GloSSAC}); details of the
construction are given in the main-text Methods. The Volcanic Aerosol
Index \autocite{VAI2003}, used as the secondary volcanic index, was
obtained from the NOAA National Centers for Environmental Information
(\url{https://www.ncei.noaa.gov/access/paleo-search/study/5786}). The
compiled index files, the raw stratospheric-aerosol NetCDF sources, and
the extracted and derived LULC panel are available in the GitHub
repository at
\href{https://github.com/everval/Airborne-Fraction-Multiple-LULC-Measurements/}{github.com/everval/Airborne-Fraction-Multiple-LULC-Measurements}.

\subsection{Reproducibility Statement}\label{reproducibility-statement}

All code used to process the data, estimate the models, and generate the
figures and tables is available in the manuscript repository
\href{https://github.com/everval/Airborne-Fraction-Multiple-LULC-Measurements/}{github.com/everval/Airborne-Fraction-Multiple-LULC-Measurements}
with DOI:
\href{https://doi.org/10.5281/zenodo.21543771}{10.5281/zenodo.21543771}.
The repository includes the Quarto sources, the Julia analysis scripts
and helper library (\texttt{scripts/}), the processed result tables
(\texttt{results/}), and a \texttt{README} with instructions for
reproducing the analysis end to end. Script execution order and the
mapping from each script to the figures and tables it produces are
documented in \texttt{REPRODUCIBILITY.md}. The code is released under
the MIT License. A preprint of this manuscript is available on
\href{https://arxiv.org/abs/2605.30242}{arXiv (2605.30242)}.

\section*{References}\label{references}
\addcontentsline{toc}{section}{References}

\printbibliography[heading=none]

@article{Betts2016,
  author  = {Betts, Richard A. and Jones, Chris D. and Knight, Jeff R.
             and Keeling, Ralph F. and Kennedy, John J.},
  title   = {El Ni\~no and a record {CO$_2$} rise},
  journal = {Nature Climate Change},
  volume  = {6},
  number  = {9},
  pages   = {806--810},
  year    = {2016},
  doi     = {10.1038/nclimate3063}
}

@article{Canadell2007,
  author  = {Canadell, J. G. and Le Qu\'er\'e, C. and Raupach, M. R.
             and Field, C. B. and Buitenhuis, E. T. and Ciais, P.
             and Conway, T. J. and Gillett, N. P.
             and Houghton, R. A. and Marland, G.},
  title   = {Contributions to accelerating atmospheric {CO$_2$} growth
             from economic activity, carbon intensity, and efficiency
             of natural sinks},
  journal = {Proceedings of the National Academy of Sciences},
  volume  = {104},
  number  = {47},
  pages   = {18866--18870},
  year    = {2007},
  doi     = {10.1073/pnas.0702737104}
}

@article{Aitken1935,
  author  = {Aitken, A. C.},
  title   = {On Least Squares and Linear Combination of Observations},
  journal = {Proceedings of the Royal Society of Edinburgh},
  volume  = {55},
  pages   = {42--48},
  year    = {1935},
  doi     = {10.1017/S0370164600014346}
}

@article{Ballantyne2012,
  author  = {Ballantyne, A. P. and Alden, C. B. and Miller, J. B.
             and Tans, P. P. and White, J. W. C.},
  title   = {Increase in observed net carbon dioxide uptake by land
             and oceans during the past 50 years},
  journal = {Nature},
  volume  = {488},
  pages   = {70--72},
  year    = {2012},
  doi     = {10.1038/nature11299}
}

@book{Carroll2006,
  author    = {Carroll, R. J. and Ruppert, D. and Stefanski, L. A. and Crainiceanu, C. M.},
  title     = {Measurement Error in Nonlinear Models: A Modern Perspective},
  edition   = {2nd},
  publisher = {Chapman \& Hall/CRC},
  address   = {Boca Raton},
  year      = {2006}
}

@book{Cochran1977,
  author    = {Cochran, W. G.},
  title     = {Sampling Techniques},
  edition   = {3rd},
  publisher = {Wiley},
  address   = {New York},
  year      = {1977}
}

@article{Friedlingstein2025,
	title = {Global carbon budget 2025},
	volume = {18},
	url = {https://essd.copernicus.org/articles/18/3211/2026/},
	doi = {10.5194/essd-18-3211-2026},
	number = {5},
	journal = {Earth System Science Data},
	author = {Friedlingstein, P. and O'Sullivan, M. and Jones, M. W. and Andrew, R. M. and Bakker, D. C. E. and Hauck, J. and Landschützer, P. and Le Quéré, C. and Li, H. and Luijkx, I. T. and Peters, G. P. and Peters, W. and Pongratz, J. and Schwingshackl, C. and Sitch, S. and Canadell, J. G. and Ciais, P. and Aas, K. and Alin, S. R. and Anthoni, P. and Barbero, L. and Bates, N. R. and Bellouin, N. and Benoit-Cattin, A. and Berghoff, C. F. and Bernardello, R. and Bopp, L. and Brasika, I. B. M. and Chamberlain, M. A. and Chandra, N. and Chevallier, F. and Chini, L. P. and Collier, N. O. and Colligan, T. H. and Cronin, M. and Djeutchouang, L. M. and Dou, X. and Enright, M. P. and Enyo, K. and Erb, M. and Evans, W. and Feely, R. A. and Feng, L. and Ford, D. J. and Foster, A. and Fransner, F. and Gasser, T. and Gehlen, M. and Gkritzalis, T. and Goncalves De Souza, J. and Grassi, G. and Gregor, L. and Gruber, N. and Guenet, B. and Gürses, Ö. and Harrington, K. and Harris, I. and Heinke, J. and Hurtt, G. C. and Iida, Y. and Ilyina, T. and Ito, A. and Jacobson, A. R. and Jain, A. K. and Jarnı́ková, T. and Jersild, A. and Jiang, F. and Jones, S. D. and Kato, E. and Keeling, R. F. and Klein Goldewijk, K. and Knauer, J. and Kong, Y. and Korsbakken, J. I. and Koven, C. and Kunimitsu, T. and Lan, X. and Liu, J. and Liu, Z. and Liu, Z. and Lo Monaco, C. and Ma, L. and Marland, G. and McGuire, P. C. and McKinley, G. A. and Melton, J. R. and Monacci, N. and Monier, E. and Morgan, E. J. and Munro, D. R. and Müller, J. D. and Nakaoka, S.-I. and Nayagam, L. R. and Niwa, Y. and Nutzel, T. and Olsen, A. and Omar, A. M. and Pan, N. and Pandey, S. and Pierrot, D. and Qin, Z. and Regnier, P. and Rehder, G. and Resplandy, L. and Roobaert, A. and Rosan, T. M. and Rödenbeck, C. and Schwinger, J. and Skjelvan, I. and Smallman, T. L. and Spada, V. and Sreeush, M. G. and Sun, Q. and Sutton, A. J. and Sweeney, C. and Swingedouw, D. and Séférian, R. and Takao, S. and Tatebe, H. and Tian, H. and Tian, X. and Tilbrook, B. and Tsujino, H. and Tubiello, F. and van Ooijen, E. and van der Werf, G. R. and van de Velde, S. J. and Walker, A. P. and Wanninkhof, R. and Yang, X. and Yuan, W. and Yue, X. and Zeng, J.},
	year = {2026},
	pages = {3211--3288},
}

@book{Fuller1987,
  author    = {Fuller, W. A.},
  title     = {Measurement Error Models},
  publisher = {Wiley},
  address   = {New York},
  year      = {1987}
}

@article{Knorr2009,
  author  = {Knorr, W.},
  title   = {Is the airborne fraction of anthropogenic {CO$_2$} emissions
             increasing?},
  journal = {Geophysical Research Letters},
  volume  = {36},
  number  = {21},
  pages   = {L21710},
  year    = {2009},
  doi     = {10.1029/2009GL040613}
}

@article{LeQuere2009,
  author  = {Le Qu\'er\'e, C. and Raupach, M. R. and Canadell, J. G.
             and Marland, G. and Bopp, L. and Ciais, P.
             and Conway, T. J. and Doney, S. C. and Feely, R. A.
             and Foster, P. and Friedlingstein, P. and Gurney, K.
             and Houghton, R. A. and House, J. I. and Huntingford, C.
             and Levy, P. E. and Lomas, M. R. and Majkut, J.
             and Metzl, N. and Ometto, J. P. and Peters, G. P.
             and Prentice, I. C. and Randerson, J. T.
             and Running, S. W. and Sarmiento, J. L. and Schuster, U.
             and Sitch, S. and Takahashi, T. and Viovy, N.
             and van der Werf, G. R. and Woodward, F. I.},
  title   = {Trends in the sources and sinks of carbon dioxide},
  journal = {Nature Geoscience},
  volume  = {2},
  pages   = {831--836},
  year    = {2009},
  doi     = {10.1038/ngeo689}
}

@article{Oehlert1992,
  author  = {Oehlert, G. W.},
  title   = {A Note on the Delta Method},
  journal = {The American Statistician},
  volume  = {46},
  number  = {1},
  pages   = {27--29},
  year    = {1992},
  doi     = {10.1080/00031305.1992.10475842}
}

@article{Raupach2007,
  author  = {Raupach, M. R. and Marland, G. and Ciais, P.
             and Le Qu\'er\'e, C. and Canadell, J. G.
             and Klepper, G. and Field, C. B.},
  title   = {Global and regional drivers of accelerating {CO$_2$}
             emissions},
  journal = {Proceedings of the National Academy of Sciences},
  volume  = {104},
  number  = {24},
  pages   = {10288--10293},
  year    = {2007},
  doi     = {10.1073/pnas.0700609104}
}

@article{veravaldes2025robustestimationco2,
      title={Robust estimation of carbon dioxide airborne fraction under measurement errors}, 
      author={J. Eduardo Vera-Valdés and Charisios Grivas},
      year={2025},
      journal={Environmental Research Communications},
      url={http://iopscience.iop.org/article/10.1088/2515-7620/adc06b}, 
      doi={10.1088/2515-7620/adc06b},
      volume={7},
      number={3},
      pages={031009},
      publisher={IOP Publishing}
}

@article{bennedsenEvidenceTrendCO22023,
	title = {On the evidence of a trend in the {CO2} airborne fraction},
	volume = {616},
	issn = {1476-4687},
	url = {https://doi.org/10.1038/s41586-023-05871-6},
	doi = {10.1038/s41586-023-05871-6},
	number = {7956},
	journal = {Nature},
	author = {Bennedsen, Mikkel and Hillebrand, Eric and Koopman, Siem Jan},
	month = apr,
	year = {2023},
	pages = {E1--E3},
}

@article{bennedsenRegressionbasedApproachCO22024,
	title = {A regression-based approach to the {CO2} airborne fraction},
	volume = {15},
	issn = {2041-1723},
	url = {https://www.nature.com/articles/s41467-024-52728-1},
	doi = {10.1038/s41467-024-52728-1},
	language = {en},
	number = {1},
	urldate = {2024-10-04},
	journal = {Nature Communications},
	author = {Bennedsen, Mikkel and Hillebrand, Eric and Koopman, Siem Jan},
	month = oct,
	year = {2024},
	pages = {8507},
}

@article{bennettQuantificationAirborneFraction2024,
	title = {Quantification of the {Airborne} {Fraction} of {Atmospheric} {CO}$_{\textrm{2}}$ {Reveals} {Stability} in {Global} {Carbon} {Sinks} {Over} the {Past} {Six} {Decades}},
	volume = {129},
	issn = {2169-8953, 2169-8961},
	url = {https://agupubs.onlinelibrary.wiley.com/doi/10.1029/2023JG007760},
	doi = {10.1029/2023JG007760},
	language = {en},
	number = {3},
	urldate = {2026-05-28},
	journal = {Journal of Geophysical Research: Biogeosciences},
	author = {Bennett, Brian F. and Salawitch, Ross J. and McBride, Laura A. and Hope, Austin P. and Tribett, Walter R.},
	month = mar,
	year = {2024},
	pages = {e2023JG007760},
}

@article{rahmstorfGlobalWarmingHas2025,
author = {Foster, G. and Rahmstorf, S.},
title = {Global Warming Has Accelerated Significantly},
journal = {Geophysical Research Letters},
volume = {53},
number = {5},
pages = {e2025GL118804},
doi = {https://doi.org/10.1029/2025GL118804},
url = {https://agupubs.onlinelibrary.wiley.com/doi/abs/10.1029/2025GL118804},
note = {e2025GL118804 2025GL118804},
year = {2026}
}

@article{Miniere2023,
  author  = {Mini{\`e}re, Aurore and von Schuckmann, Karina and Sall{\'e}e, Jean-Baptiste and Vogt, Lukas},
  title   = {Robust acceleration of Earth system heating observed over the past six decades},
  journal = {Scientific Reports},
  volume  = {13},
  pages   = {22975},
  year    = {2023},
  doi     = {10.1038/s41598-023-49353-1}
}

@article{StortoYang2024,
  author  = {Storto, Andrea and Yang, Chunxue},
  title   = {Acceleration of the ocean warming from 1961 to 2022 unveiled by large-ensemble reanalyses},
  journal = {Nature Communications},
  volume  = {15},
  pages   = {545},
  year    = {2024},
  doi     = {10.1038/s41467-024-44749-7}
}

@misc{WMO_SGC_2025,
  author       = {{World Meteorological Organization}},
  title        = {State of the Global Climate 2025},
  year         = {2026},
  howpublished = {Report WMO-No. 1391},
  institution  = {World Meteorological Organization (WMO)},
  address      = {Geneva},
  doi          = {10.59327/WMO/S/CRI/SOC/1},
  url          = {https://wmo.int/publication-series/state-of-global-climate/state-of-global-climate-2025},
  note         = {ISBN 978-92-63-11391-7},
}

@misc{Lan2025,
  author       = {Lan, Xin and Tans, Pieter and Thoning, K. W.},
  title        = {Trends in globally-averaged CO2 determined from NOAA Global Monitoring Laboratory measurements},
  year         = {2026},
  note         = {Version Tuesday, 05-May-2026 05:42:49 MDT},
  url          = {https://gml.noaa.gov/ccgg/trends/global.html},
  doi          = {10.15138/9N0H-ZH07},
  institution  = {NOAA Global Monitoring Laboratory}
}

@article{Hansis2015,
  author  = {Hansis, E. and Davis, S. J. and Pongratz, J.},
  title   = {Relevance of methodological choices for accounting of land use change carbon fluxes},
  journal = {Global Biogeochemical Cycles},
  volume  = {29},
  pages   = {1230--1246},
  year    = {2015},
  doi     = {10.1002/2014GB004997}
}

@article{Gasser2020,
  author  = {Gasser, T. and Crepin, L. and Quilcaille, Y. and Houghton, R. A. and Ciais, P. and Obersteiner, M.},
  title   = {Historical {CO2} emissions from land use and land cover change and their uncertainty},
  journal = {Biogeosciences},
  volume  = {17},
  pages   = {4075--4101},
  year    = {2020},
  doi     = {10.5194/bg-17-4075-2020}
}

@article{Qin2024,
  author  = {Qin, Z. and Zhu, Y. and Canadell, J. G. and Chen, M. and Li, T. and Mishra, U. and Yuan, W.},
  title   = {Global spatially explicit carbon emissions from land-use change over the past six decades (1961-2020)},
  journal = {One Earth},
  volume  = {7},
  pages   = {835--847},
  year    = {2024},
  doi     = {10.1016/j.oneear.2024.04.002}
}

@article{Conchedda2020,
  author  = {Conchedda, G. and Tubiello, F. N.},
  title   = {Drainage of organic soils and GHG emissions: validation with country data},
  journal = {Earth System Science Data},
  volume  = {12},
  pages   = {3113--3137},
  year    = {2020},
  doi     = {10.5194/essd-12-3113-2020},
  url     = {https://essd.copernicus.org/articles/12/3113/2020/}
}

@article{Mueller2021,
  author  = {M{\"u}ller, J. and Joos, F.},
  title   = {Committed and projected future changes in global peatlands -- continued transient model simulations since the Last Glacial Maximum},
  journal = {Biogeosciences},
  volume  = {18},
  pages   = {3657--3687},
  year    = {2021},
  doi     = {10.5194/bg-18-3657-2021},
  url     = {https://bg.copernicus.org/articles/18/3657/2021/}
}

@article{Qiu2021,
  author  = {Qiu, C. and Ciais, P. and Zhu, D. and Guenet, B. and Peng, S. and Petrescu, A. M. R. and Lauerwald, R. and Makowski, D. and Gallego-Sala, A. V. and Charman, D. J. and Brewer, S. C.},
  title   = {Large historical carbon emissions from cultivated northern peatlands},
  journal = {Science Advances},
  volume  = {7},
  pages   = {eabf1332},
  year    = {2021},
  doi     = {10.1126/sciadv.abf1332},
  url     = {https://www.science.org/doi/10.1126/sciadv.abf1332}
}

@article{Haverd2018,
  author  = {Haverd, V. and Smith, B. and Nieradzik, L. and Briggs, P. R. and Woodgate, W. and Trudinger, C. M. and Canadell, J. G. and Cuntz, M.},
  title   = {A new version of the CABLE land surface model (Subversion revision r4601) incorporating land use and land cover change, woody vegetation demography, and a novel optimisation-based approach to plant coordination of photosynthesis},
  journal = {Geoscientific Model Development},
  volume  = {11},
  pages   = {2995--3026},
  year    = {2018},
  doi     = {10.5194/gmd-11-2995-2018},
  url     = {https://gmd.copernicus.org/articles/11/2995/2018/}
}

@article{Melton2020,
  author  = {Melton, J. R. and Arora, V. K. and Wisernig-Cojoc, E. and Seiler, C. and Fortier, M. and Chan, E.},
  title   = {CLASSIC v1.0: the open-source community successor to the Canadian Land Surface Scheme (CLASS) and the Canadian Terrestrial Ecosystem Model (CTEM) -- Part 1: Model framework and site-level performance},
  journal = {Geoscientific Model Development},
  volume  = {13},
  pages   = {2825--2850},
  year    = {2020},
  doi     = {10.5194/gmd-13-2825-2020}
}

@article{Lawrence2019,
  author  = {Lawrence, D. M. and Fisher, R. A. and Koven, C. D. and Oleson, K. W. and Swenson, S. C. and Bonan, G. and others},
  title   = {The Community Land Model version 5: Description of new features, benchmarking, and impact of forcing uncertainty},
  journal = {Journal of Advances in Modeling Earth Systems},
  volume  = {11},
  pages   = {4245--4287},
  year    = {2019},
  doi     = {10.1029/2018MS001583}
}

@article{Fisher2015,
  author  = {Fisher, R. A. and Muszala, S. and Verteinstein, M. and Lawrence, P. and Xu, C. and McDowell, N. G. and Knox, R. G. and Koven, C. and Holm, J. and Rogers, B. M. and Spessa, A. and Lawrence, D. and Bonan, G.},
  title   = {Taking off the training wheels: the properties of a dynamic vegetation model without climate envelopes, CLM4.5(ED)},
  journal = {Geoscientific Model Development},
  volume  = {8},
  pages   = {3593--3619},
  year    = {2015},
  doi     = {10.5194/gmd-8-3593-2015}
}

@article{Tian2015,
  author  = {Tian, H. and Chen, G. and Lu, C. and Xu, X. and Hayes, D. J. and Ren, W. and Pan, S. and Huntzinger, D. N. and Wofsy, S. C.},
  title   = {North American terrestrial CO2 uptake largely offset by CH4 and N2O emissions: Toward a full accounting of the greenhouse gas budget},
  journal = {Climatic Change},
  volume  = {129},
  pages   = {423--426},
  year    = {2015}
}

@article{Ma2022,
  author  = {Ma, L. and Hurtt, G. and Ott, L. and Sahajpal, R. and Fisk, J. and Lamb, R. and Tang, H. and Flanagan, S. and Chini, L. and Chatterjee, A. and Sullivan, J.},
  title   = {Global evaluation of the Ecosystem Demography model (ED v3.0)},
  journal = {Geoscientific Model Development},
  volume   = {15},
  pages    = {1971--1994},
  year     = {2022},
  doi      = {10.5194/gmd-15-1971-2022},
  url      = {https://gmd.copernicus.org/articles/15/1971/2022/}
}

@article{Yang2023,
  author  = {Yang, X. and Thornton, P. and Ricciuto, D. and Wang, Y. and Hoffman, F.},
  title   = {Global evaluation of terrestrial biogeochemistry in the Energy Exascale Earth System Model (E3SM) and the role of the phosphorus cycle in the historical terrestrial carbon balance},
  journal = {Biogeosciences},
  volume   = {20},
  pages    = {2813--2836},
  year     = {2023},
  doi      = {10.5194/bg-20-2813-2023},
  url      = {https://bg.copernicus.org/articles/20/2813/2023/}
}

@article{Needham2025,
  author  = {Needham, J. and others},
  title   = {Vertical canopy gradients of respiration drive plant carbon budgets and leaf area index},
  journal = {New Phytologist},
  volume   = {246},
  pages    = {144--157},
  year     = {2025},
  doi      = {10.1111/nph.20423}
}

@article{Felzer2018,
  author  = {Felzer, B. S. and Jiang, M.},
  title   = {Effect of land use and land cover change in context of growth enhancements in the United States since 1700: Net source or sink?},
  journal = {Journal of Geophysical Research: Biogeosciences},
  volume   = {123},
  pages    = {3439--3457},
  year     = {2018},
  doi      = {10.1029/2017JG004378}
}

@article{Xia2024,
  author  = {Xia, J. Z. and Ren, P. Y. and Wang, X. H. and Liu, D. and Chen, X. Z. and Dan, L. and He, B. and He, H. L. and Ju, W. M. and Liang, M. Q. and Lu, X. J. and Peng, J. and Qin, Z. C. and Xia, J. Z. and Zheng, B. and Wei, J. and Yue, X. and Yu, G. R. and Piao, S. L. and Yuan, W. P.},
  title   = {The carbon budget of China: 1980--2021},
  journal = {Science Bulletin},
  volume   = {69},
  pages    = {114--124},
  year     = {2024}
}

@article{Yue2024,
  author  = {Yue, X. and Zhou, H. and Tian, C. and Ma, Y. and Hu, Y. and Gong, C. and Zheng, H. and Liao, H.},
  title   = {Development and evaluation of the interactive Model for Air Pollution and Land Ecosystems (iMAPLE) version 1.0},
  journal = {Geoscientific Model Development},
  volume   = {17},
  pages    = {4621--4642},
  year     = {2024}
}

@article{Shu2020,
  author  = {Shu, S. and Jain, A. K. and Koven, C. D. and Mishra, U.},
  title   = {Estimation of Permafrost SOC Stock and Turnover Time Using a Land Surface Model With Vertical Heterogeneity of Permafrost Soils},
  journal = {Global Biogeochemical Cycles},
  volume   = {34},
  pages   = {e2020GB006585},
  year    = {2020},
  doi     = {10.1029/2020GB006585}
}

@misc{Reick2021,
  author = {Reick, C. H. and Gayler, V. and Goll, D. and Hagemann, S. and Heidkamp, M. and Nabel, J. E. M. S. and Raddatz, T. and Roeckner, E. and Schnur, R. and Wilkenskjeld, S.},
  title  = {JSBACH 3 - The land component of the MPI Earth System Model: documentation of version 3.2},
  year   = {2021},
  doi    = {10.17617/2.3279802}
}

@article{Poulter2011,
  author  = {Poulter, B. and Frank, D. C. and Hodson, E. L. and Zimmermann, N. E.},
  title   = {Impacts of land cover and climate data selection on understanding terrestrial carbon dynamics and the CO2 airborne fraction},
  journal = {Biogeosciences},
  volume   = {8},
  pages    = {2027--2036},
  year     = {2011}
}

@article{Smith2014,
  author  = {Smith, B. and Warlind, D. and Arneth, A. and Hickler, T. and Leadley, P. and Siltberg, J. and Zaehle, S.},
  title   = {Implications of incorporating N cycling and N limitations on primary production in an individual-based dynamic vegetation model},
  journal = {Biogeosciences},
  volume   = {11},
  pages    = {2027--2054},
  year     = {2014}
}

@article{Schaphoff2018,
  author  = {Schaphoff, S. and von Bloh, W. and Rammig, A. and Thonicke, K. and Biemans, H. and Forkel, M. and Gerten, D. and Heinke, J. and J{\"a}germeyr, J. and Knauer, J. and Langerwisch, F. and Lucht, W. and M{\"u}ller, C. and Rolinski, S. and Waha, K.},
  title   = {LPJmL4 -- a dynamic global vegetation model with managed land -- Part 1: Model description},
  journal = {Geoscientific Model Development},
  volume   = {11},
  pages    = {1343--1375},
  year     = {2018},
  doi      = {10.5194/gmd-11-1343-2018}
}

@article{Lienert2018,
  author  = {Lienert, S. and Joos, F.},
  title   = {A Bayesian ensemble data assimilation to constrain model parameters and land-use carbon emissions},
  journal = {Biogeosciences},
  volume   = {15},
  pages    = {2909--2930},
  year     = {2018}
}

@article{Vuichard2019,
  author  = {Vuichard, N. and Messina, P. and Luyssaert, S. and Guenet, B. and Zaehle, S. and Ghattas, J. and Ipsl, L. and Paris-Saclay, U.},
  title   = {Accounting for carbon and nitrogen interactions in the global terrestrial ecosystem model ORCHIDEE (trunk version, rev 4999): multi-scale evaluation of gross primary production},
  journal = {Geoscientific Model Development},
  volume   = {12},
  pages   = {4751--4779},
  year     = {2019}
}

@article{Walker2017,
  author  = {Walker, A. P. and Quaife, T. and van Bodegom, P. M. and De Kauwe, M. G. and Keenan, T. F. and Joiner, J. and Lomas, M. R. and MacBean, N. and Xu, C. G. and Yang, X. J. and Woodward, F. I.},
  title   = {The impact of alternative trait-scaling hypotheses for the maximum photosynthetic carboxylation rate (V-cmax) on global gross primary production},
  journal = {New Phytologist},
  volume   = {215},
  pages   = {1370--1386},
  year    = {2017}
}

@article{Kato2013,
  author  = {Kato, E. and Kinoshita, T. and Ito, A. and Kawamiya, M. and Yamagata, Y.},
  title   = {Evaluation of spatially explicit emission scenario of land-use change and biomass burning using a process-based biogeochemical model},
  journal = {Journal of Land Use Science},
  volume   = {8},
  number   = {1},
  pages   = {104--122},
  year     = {2013},
  doi      = {10.1080/1747423X.2011.628705}
}

@article{Ito2019,
  author  = {Ito, A.},
  title   = {Disequilibrium of terrestrial ecosystem CO2 budget caused by disturbance-induced emissions and non-CO2 carbon export flows: a global model assessment},
  journal = {Earth System Dynamics},
  volume   = {10},
  pages    = {685--709},
  year     = {2019},
  doi      = {10.5194/esd-10-685-2019}
}

@article{Henderson1953,
  author  = {Henderson, Charles R.},
  title   = {Estimation of Variance and Covariance Components},
  journal = {Biometrics},
  year    = {1953},
  volume  = {9},
  number  = {2},
  pages   = {226--252},
  doi     = {10.2307/3001881}
}

@article{Bolker2009,
  author  = {Bolker, Benjamin M. and Brooks, Mollie E. and Clark, Connie J. and Geange, Shane W. and Poulsen, John R. and Stevens, M. Henry H. and White, J. -S.},
  title   = {Generalized linear mixed models: a practical guide for ecology and evolution},
  journal = {Trends in Ecology \& Evolution},
  year    = {2009},
  volume  = {24},
  number  = {3},
  pages   = {127--135},
  doi     = {10.1016/j.tree.2008.10.008}
}

@article{Gloor2010CarbonFeedbackAF,
  author  = {Manfred Gloor and others},
  title   = {What Can Be Learned about Carbon Cycle Climate Feedbacks from the CO2 Airborne Fraction?},
  journal = {Atmospheric Chemistry and Physics},
  year    = {2010},
  volume  = {10},
  pages   = {7739--7752},
  url     = {https://acp.copernicus.org/articles/10/7739/2010/}
}

@article{LHeureux2024RONI,
  author = "Michelle L. L’Heureux and Michael K. Tippett and Matthew C. Wheeler and Hanh Nguyen and Sugata Narsey and Nathaniel Johnson and Zeng-Zhen Hu and Andrew B. Watkins and Chris Lucas and Catherine Ganter and Emily Becker and Wanqiu Wang and Tom Di Liberto",
  title = "A Relative Sea Surface Temperature Index for Classifying ENSO Events in a Changing Climate",
  journal = "Journal of Climate",
  year = "2024",
  publisher = "American Meteorological Society",
  volume = "37",
  number = "4",
  doi = "10.1175/JCLI-D-23-0406.1",
  pages=      "1197 - 1211",
  url = "https://journals.ametsoc.org/view/journals/clim/37/4/JCLI-D-23-0406.1.xml"
}

@article{Huang_2017_ERSSTv5_JCLI,
  author  = {Huang, Boyin and Thorne, Peter W. and Banzon, Viva F. 
             and Boyer, Tim and Chepurin, Gennady and Lawrimore, Jay H. 
             and Menne, Matthew J. and Smith, Thomas M. 
             and Vose, Russell S. and Zhang, Huai-Min},
  title   = {Extended Reconstructed Sea Surface Temperature, Version 5 (ERSSTv5): 
             Upgrades, Validations, and Intercomparisons},
  journal = {Journal of Climate},
  year    = {2017},
  doi     = {10.1175/JCLI-D-16-0836.1}
}

@article{VAI2003,
author = {Ammann, Caspar M. and Meehl, Gerald A. and Washington, Warren M. and Zender, Charles S.},
title = {A monthly and latitudinally varying volcanic forcing dataset in simulations of 20th century climate},
journal = {Geophysical Research Letters},
volume = {30},
number = {12},
pages = {},
doi = {10.1029/2003GL016875},
url = {https://agupubs.onlinelibrary.wiley.com/doi/abs/10.1029/2003GL016875},
eprint = {https://agupubs.onlinelibrary.wiley.com/doi/pdf/10.1029/2003GL016875},
year = {2003}
}

@book{IPCC_AR6_WG1,
  author    = {{IPCC}},
  title     = {Climate Change 2021: The Physical Science Basis},
  year      = {2021},
  editor    = {Masson-Delmotte, Val{\'e}rie and Zhai, Panmao and Pirani, Anna and Connors, Sarah L. and P{\'e}an, Clotilde and Berger, Sophie and Caud, Nathalie and Chen, Yand and Goldfarb, Leah and Gomis, Melissa I. and Huang, Muyin and Legrand, Richard and Lixin, Tian and Marotzke, J{\"u}rgen and Naik, Vaishali and Seneviratne, Sonia I. and Slade, Raphael and Turpali, Clara and Zhou, Bailang},
  series    = {Contribution of Working Group I to the Sixth Assessment Report of the Intergovernmental Panel on Climate Change},
  publisher = {Cambridge University Press},
  doi       = {10.1017/9781009157896}
}

@article{Aubry2025CMIP7,
  author  = {Aubry, Thomas J. and Toohey, Matthew and Marshall, Lauren and Schmidt, Anja and Rougier, Jonathan and others},
  title   = {Stratospheric aerosol forcing for {CMIP7}: optical properties for pre-industrial, historical, and scenario simulations},
  journal = {Geoscientific Model Development},
  volume  = {19},
  pages   = {3725--3760},
  year    = {2026},
  doi     = {10.5194/gmd-19-3725-2026},
  note    = {CMIP7 stratospheric aerosol forcing v2.2.1, University of Exeter, input4MIPs dataset UOEXETER-CMIP-2-2-1}
}

@dataset{Aubry2025CMIP7data,
  author    = {Aubry, Thomas J. and Toohey, Matthew and Marshall, Lauren and Schmidt, Anja and others},
  title     = {{CMIP7} historical stratospheric aerosol optical properties and stratospheric volcanic sulfur emissions, version 2.2.1},
  year      = {2025},
  publisher = {input4MIPs, Earth System Grid Federation},
  version   = {UOEXETER-CMIP-2-2-1},
  doi       = {10.25981/ESGF.input4MIPs.CMIP7/2522673},
  url       = {https://aims2.llnl.gov/search/input4MIPs/}
}

@article{Kovilakam2020GloSSAC,
  author  = {Kovilakam, Mahesh and Thomason, Larry W. and Ernest, Nicholas and Rieger, Landon and Bourassa, Adam and Millán, Luis},
  title   = {The {Global} {Space-based} {Stratospheric} {Aerosol} {Climatology} ({GloSSAC}) version 2.0: 1979--2018},
  journal = {Earth System Science Data},
  volume  = {12},
  number  = {4},
  pages   = {2607--2634},
  year    = {2020},
  doi     = {10.5194/essd-12-2607-2020},
  note    = {Data updated to GloSSAC v2.24 (1979--2025); NASA Atmospheric Science Data Center}
}

@dataset{Kovilakam2026GloSSACdata,
  author    = {Kovilakam, Mahesh},
  title     = {Global Space-based Stratospheric Aerosol Climatology Version 2.24},
  year      = {2026},
  publisher = {NASA Langley Atmospheric Science Data Center Distributed Active Archive Center (ASDC DAAC)},
  version   = {2.24},
  doi       = {10.5067/GLOSSAC-L3-V2.24},
  url       = {https://www.earthdata.nasa.gov/data/catalog/larc-cloud-glossac-2.24}
}

@article{Smith2000,
  author  = {Smith, Catherine A. and Sardeshmukh, Prashant D.},
  title   = {The effect of {ENSO} on the intraseasonal variance of surface temperature in winter},
  journal = {International Journal of Climatology},
  volume  = {20},
  number  = {13},
  pages   = {1543--1557},
  year    = {2000},
  doi     = {10.1002/1097-0088(20001115)20:13<1543::AID-JOC579>3.0.CO;2-A}
}

@misc{NOAA_PSL_ENSO,
  author       = {{NOAA Physical Sciences Laboratory}},
  title        = {El Ni{\~n}o/Southern Oscillation ({ENSO}) indices},
  howpublished = {\url{https://psl.noaa.gov/enso/}},
  year         = {2025},
  note         = {NOAA PSL, Boulder, Colorado, USA}
}

@article{Andrews1991,
  author  = {Andrews, Donald W. K.},
  title   = {Heteroskedasticity and Autocorrelation Consistent Covariance Matrix Estimation},
  journal = {Econometrica},
  volume  = {59},
  number  = {3},
  pages   = {817--858},
  year    = {1991},
  doi     = {10.2307/2938229}
}

@article{NeweyWest1987,
  author  = {Newey, Whitney K. and West, Kenneth D.},
  title   = {A Simple, Positive Semi-Definite, Heteroskedasticity and Autocorrelation Consistent Covariance Matrix},
  journal = {Econometrica},
  volume  = {55},
  number  = {3},
  pages   = {703--708},
  year    = {1987},
  doi     = {10.2307/1913610}
}

@techreport{PraisWinsten1954,
  author      = {Prais, Sigbert J. and Winsten, Christopher B.},
  title       = {Trend Estimators and Serial Correlation},
  institution = {Cowles Commission for Research in Economics, University of Chicago},
  type        = {Cowles Commission Discussion Paper, Statistics},
  number      = {383},
  year        = {1954}
}

@article{VeraValdes2026acceleration,
  title   = {Global Warming Has Been Accelerating Since At Least 1990},
  author  = {Vera-Vald{\'e}s, J. Eduardo},
  journal = {arXiv (2606.04114)},
  year    = {2026},
  doi     = {10.48550/arXiv.2606.04114},
  url     = {https://arxiv.org/abs/2606.04114}
}

@article{bugajski2026spatialemergenceaccelerationglobal,
      title={Spatial emergence of acceleration in global warming}, 
      author={Tanja Korsten Bugajski and Nicolai Peder Bulow Pedersen and J. Eduardo Vera-Valdes},
      journal={arXiv preprint arXiv:2606.18806},
      year={2026},
      doi={10.48550/arXiv.2606.18806},
      url={https://arxiv.org/abs/2606.18806}, 
}

\end{document}